\documentclass[a4paper,fleqn,usenatbib]{mnras}

\usepackage{newtxtext,newtxmath}

\usepackage[T1]{fontenc}
\usepackage{ae,aecompl}

\usepackage{graphicx}
\usepackage{natbib}
\usepackage{capt-of}
\usepackage{textcomp}
\usepackage{epstopdf}
\usepackage{amsmath,scalerel}
\usepackage{multirow}
\usepackage{url}
\usepackage{rotating}
\bibpunct{(}{)}{,}{a}{}{,}

\usepackage{graphicx}	
\usepackage{amsmath}	
\usepackage{amssymb}	

\title[Observations and laboratory work on CH$_{3}$NCO]{The ALMA-PILS survey: Detection of CH$_{3}$NCO toward the low-mass protostar IRAS 16293-2422 and laboratory constraints on its formation}

\author[N. F. W. Ligterink et al.]{
N. F. W. Ligterink,$^{1,2}$\thanks{E-mail: ligterink@strw.leidenuniv.nl}
A. Coutens,$^{3}$\thanks{E-mail: a.coutens@ucl.ac.uk}
V. Kofman,$^{1}$
H. S. P. M\"{u}ller,$^{4}$
R. T. Garrod,$^{5}$
\newauthor
H. Calcutt,$^{6}$
S. F. Wampfler,$^{7}$
J. K. J\o rgensen,$^{6}$       
H. Linnartz,$^{1}$
E. F. van Dishoeck$^{2,8}$
\\
$^{1}$Sackler Laboratory for Astrophysics, Leiden Observatory, Leiden University, PO Box 9513, 2300 RA Leiden, The Netherlands\\      
$^{2}$Leiden Observatory, Leiden University, PO Box 9513, 2300 RA Leiden, The Netherlands\\
$^{3}$Department of Physics and Astronomy, University College London, Gower St., London, WC1E 6BT, UK\\
$^{4}$I. Physikalisches Institut, Universit\"{a}t zu K\"{o}ln, Z\"{u}lpicher Str. 77, 50937 K\"{o}ln, Germany\\ 
$^{5}$Departments of Chemistry and Astronomy, University of Virginia, Charlottesville, VA 22904, USA\\
$^{6}$Centre for Star and Planet Formation, Niels Bohr Institute \& Natural History Museum of Denmark, University of Copenhagen, \O ster Voldgade 5-7, 1350 Copenhagen K., Denmark\\
$^{7}$Center for Space and Habitability, University of Bern, Sidlerstrasse 5, CH-3012 Bern, Switzerland\\
$^{8}$ Max-Planck Institut f\"{u}r Extraterrestrische Physik (MPE), Giessenbachstr. 1, 85748 Garching, Germany\\
}

\date{Accepted 2017 April 7. Received 2017 March 26; in original form 2017 January 24}

\pubyear{2017}

\begin{document}
\label{firstpage}
\pagerange{\pageref{firstpage}--\pageref{lastpage}}
\maketitle

\begin{abstract}

Methyl isocyanate (CH$_{3}$NCO) belongs to a select group of
interstellar molecules considered to be relevant precursors in the
formation of larger organic compounds, including those with peptide
bonds. The molecule has only been detected in a couple of high-mass
protostars and potentially on comets. A formation route on icy grains has been postulated for this molecule but experimental evidence is lacking. Here we extend the range of environments where methyl isocyanate is found, and unambiguously identify CH$_{3}$NCO through the detection of 43 unblended transitions in the ALMA Protostellar Interferometric Line Survey (PILS) of the low mass 
solar-type protostellar binary IRAS 16293-2422. The molecule is detected toward
both components of the binary with a ratio HNCO/CH$_3$NCO
$\sim$4--12. The isomers CH$_{3}$CNO and CH$_3$OCN are not identified,
resulting in upper abundance ratios of CH$_{3}$NCO/CH$_{3}$CNO > 100
and CH$_{3}$NCO/CH$_3$OCN > 10. The resulting abundance ratios compare
well with those found for related N-containing species toward
high-mass protostars.  To constrain its formation, a set of cryogenic
UHV experiments is performed.  VUV irradiation of CH$_{4}$:HNCO
mixtures at 20 K strongly indicate that methyl isocyanate can be
formed in the solid-state through CH$_{3}$ and (H)NCO
recombinations. Combined with gas-grain models that include this
reaction, the solid-state route is found to be a plausible scenario to
explain the methyl isocyanate abundances found in IRAS 16293-2422.

\end{abstract}




\section{Introduction}
\label{sec.int}

\begin{figure}
\begin{center}
\includegraphics[width=\hsize]{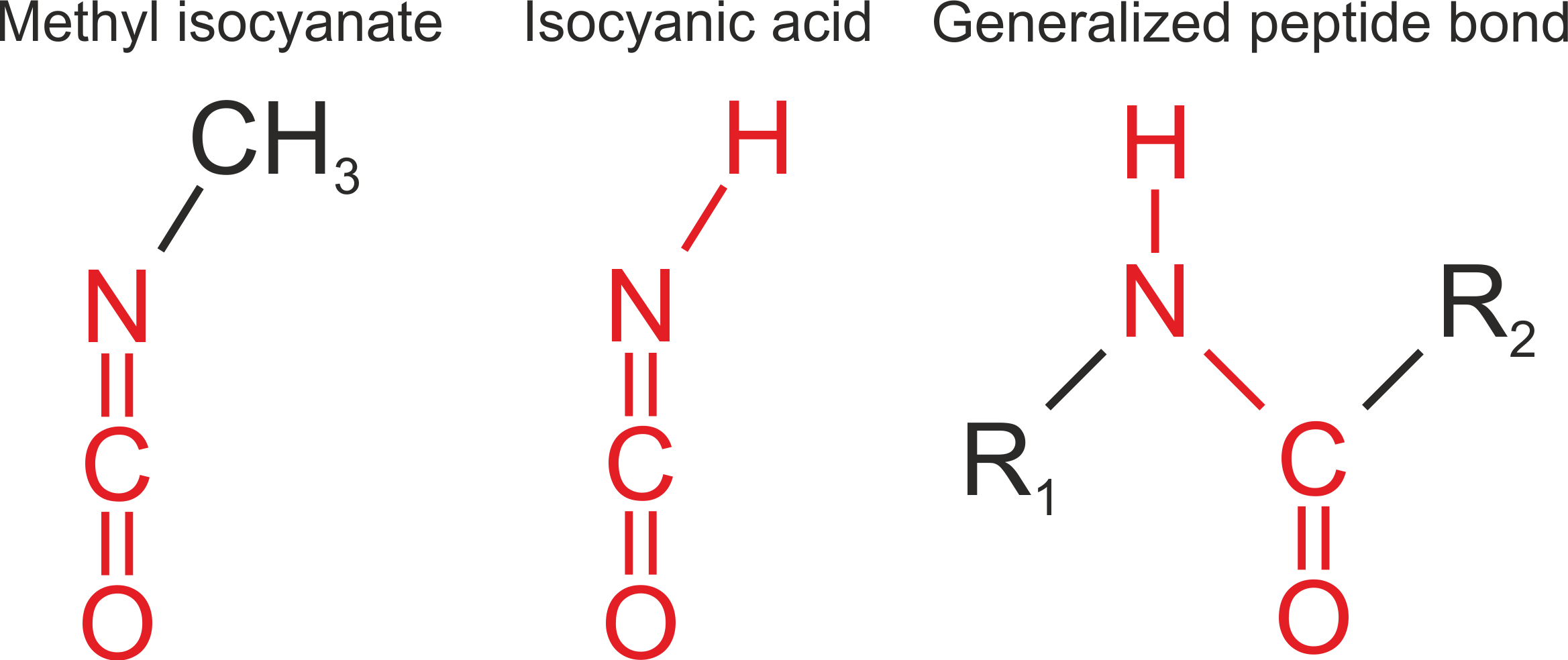}
\caption{Structure of methyl isocyanate (left), isocyanic acid (middle) and the generalized structure of a peptide bond (right). In each structure the components of the peptide bond are highlighted in red. R$_{1}$ and R$_{2}$ are different molecular functional groups, which can, among other possibilities, be a methyl group as for methyl isocyanate.}
\label{fig.structure}
\end{center}
\end{figure}

Complex organic molecules, defined in astrochemistry as molecules that
consist of six or more atoms of which at least one is a carbon atom,
are widely found in star-forming regions \citep{herbstdishoeck2009}. A special category of complex molecules is that of the prebiotics, molecules that can be linked via their chemical structures or reactivity to life bearing molecules, such as amino-acids and sugars. Methyl isocyanate, CH$_{3}$NCO, also
known as isocyanomethane, is a molecule that falls in this category,
because of its structural similarity with a peptide bond (Figure
\ref{fig.structure}). This type of bond connects amino-acids to form
proteins and as such is interesting because it connects to chemistry relevant to the formation of the building blocks of life. 

The majority of identified complex molecules has mainly been detected toward
high-mass hot cores, such as Orion KL and Sgr B2
\citep[e.g.,][]{blake1987,nummelin2000,belloche2013,belloche2014,tercero2013,neill2014,crockett2014},
but over the past decades detections toward low mass, sun-like,
protostars such as IRAS 16293-2422 (hereafter IRAS16293) have been
regularly reported as well. IRAS16293 (d = 120 pc) is considered as a protostellar template
for low-mass sources and is particularly rich in organic
molecules \citep{vandishoeck1995,cazaux2003,bottinelli2004,kuan2004,bisschop2008,jaber2014}. \citet{jorgensen2012} demonstrated the capabilities of the Atacama Large Millimeter/submillimeter Array (ALMA) with the detection of the prebiotic molecule glycolaldehyde \citep[see also][for a history of chemical studies of this source]{jorgensen2016}. More recently, other complex molecules (acetone, propanal, ethylene oxide) were identified in the framework of the PILS survey \citep{lykke2017}. Even the deuterated isotopologues of several complex molecules were detected toward this source \citep{parise2003,coutens2016,jorgensen2016}.

Unlike other molecules such as isocyanic acid (HNCO) and formamide
(NH$_{2}$CHO)\citep{bisschop2007,lopez-sepulcre2015,coutens2016},
which have a similar peptide-like structure as methyl isocyanate,
CH$_{3}$NCO only recently started to attract attention in the
astrochemistry community. This activity was inspired by a report from
\citet{goesmann2015} that CH$_3$NCO may be abundantly present on the
surface of comet 67P/Churyumov-Gerasimenko, as measured with the {\it
  Cometary Sampler and Composition} (COSAC) instrument of Rosetta's
Philae lander. Its measured high abundance of 1.3\% with respect to
H$_2$O and CH$_3$NCO/HNCO=4.3 was based on the assumption that the mass peak recorded at
$m/z=57$ is dominated by CH$_3$NCO, as COSAC's low mass resolution does
not allow discrimination of different species with nearly identical mass. Recent
measurements with the {\it Rosetta Orbiter Spectrometer for Ion and
  Neutral Analysis} (ROSINA) at much higher mass resolution showed
that the CH$_3$NCO coma abundance is significantly lower (K.\ Altwegg,
private communication). Nevertheless, the possibility of linking
complex molecules in star- and planet-forming regions with those
present in comets triggered the search for methyl isocyanate in the
interstellar medium.

Interstellar CH$_3$NCO was first detected toward Sgr B2(N) by
\citet{halfen2015} at low rotational temperatures of $\sim$25 K with a
column density ratio of $N$(HNCO)/$N$(CH$_{3}$NCO) = 35--53 depending
on the specific velocity component. \citet{cernicharo2016} detected
methyl isocyanate toward Orion KL at $N$(HNCO)/$N$(CH$_{3}$NCO)
$\simeq$ 15 and $T_{\rm ex}$ = 150 K. Their observations toward the
cold prestellar core B1-b did not yield a detection of the molecule
down to a upper column density limit of $< 2 \times$ 10$^{11}$
cm$^{-2}$ or $N$(HNCO)/$N$(CH$_{3}$NCO) > 42 \citep[based on a HNCO detection toward the same source by][]{lopez-sepulcre2015}. In the same
paper, publically accessible Sgr B2 observations from \citet{belloche2013} were
re-analysed with an updated spectral line list and, interestingly,
yielded a detection of warm methyl isocyanate at $T_{\rm ex} \simeq$
200 K and $N$(HNCO)/$N$(CH$_{3}$NCO) $\simeq$ 40.  Attempts to also
identify the methyl isocyanate isomer CH$_{3}$CNO were unsuccessful
down to $N$(CH$_{3}$NCO)/$N$(CH$_{3}$CNO) $> 100$.

The astrochemical origin of methyl isocyanate is not yet understood
and this is partly due to the limited number of laboratory studies
that have been performed. \citet{hendersongudipati2015} tentatively
assigned a mass fragmentation peak to CH$_{3}$NCO after VUV
irradiating solid-state mixtures of NH$_{3}$:CH$_{3}$OH. In other
experiments by \citet{ruzi2012} UV irradiation of frozen
$n$-methylformamide (CH$_{3}$NHCHO) also yielded methyl isocyanate,
although it was concluded to represent a minor product channel.

A number of formation routes have been hypothesized by
astrochemists. \citet{halfen2015} postulated gas-phase formation by
HNCO or HOCN methylation:

\begin{equation}
\label{eq.halfen1}
	\rm
	HNCO/HOCN(g) + CH_{3}(g) \rightarrow CH_{3}NCO(g) + H(g)
\end{equation}

\noindent
or reactions of HNCO or HOCN with protonated methane, followed by electron recombination:

\begin{equation}
\label{eq.halfen2}
	\rm
	HNCO/HOCN(g) + CH_{5}^{+}(g) \rightarrow CH_{3}NCOH^{+}(g) + H_{2}(g) 
\end{equation}
\begin{equation}
\label{eq.halfen3}
	\rm
	CH_{3}NCOH^{+}(g) + e^{-} \rightarrow CH_{3}NCO(g) + H(g)
\end{equation}

\citet{cernicharo2016} favoured solid-state formation mechanisms based
on the detection of CH$_{3}$NCO toward hot cores and its non-detection
in the cold dark cloud B1-b. Particularly, the methylation of HNCO has been
mentioned as a possible route to form methyl isocyanate in the solid-state, i.e., on the surface of an icy dust grain:

\begin{equation}
\label{eq.cernicharo1}
	\rm
	HNCO(s) + CH_{3}(s) \rightarrow CH_{3}NCO(s) + H(s)
\end{equation}

\citet{belloche2017} used the grain-surface radical-addition reaction
CH$_3$ + NCO $\rightarrow$ CH$_{3}$NCO in their models, with most of
the NCO formed via H-abstraction of HNCO:

\begin{equation}
\label{eq.belloche1}
	\rm
	HNCO(s) + H(s) \rightarrow NCO(s) + H_{2}(g)
\end{equation}
\begin{equation}
\label{eq.belloche2}
	\rm
	NCO(s) + CH_{3}(g,s) \rightarrow CH_{3}NCO(s) 
\end{equation}

These postulated routes require the reactants to be present in
sufficient amounts. Gaseous HNCO is detected in high abundances in
protostellar environments and has been imaged in IRAS16293, showing it
to be prominent in both source A and B
\citep{bisschop2008,coutens2016}. It likely results from sublimation
of OCN$^-$, known to be a major ice component in low-mass protostellar
envelopes \citep{vanbroekhuizen2005}. A significant abundance of
CH$_3$ gas is a more speculative assumption since the molecule can only be observed
by infrared spectroscopy and has so far only been seen in diffuse gas
toward the Galactic Center \citep{feuchtgruber2000}. Alternatively,
CH$_3$ radicals can be produced {\it in situ} in ices by
photodissociation of known abundant ice components like CH$_4$ or
CH$_3$OH and then react with HNCO or OCN$^-$. This is the solid-state
route that is investigated here.

In this work we present the first detection of methyl isocyanate
toward both components of the low mass protobinary IRAS16293 on scales
of $< 100$ AU. An independent detection toward source B is also reported by \citet{martin-domenech2017}. A set of systematic laboratory experiments is presented in order to validate the solid-state formation routes of
CH$_{3}$NCO. The observational work is presented in Section
\ref{sec.obs} and the laboratory work in Section
\ref{sec.lab}. In Section \ref{sec.dis} the results of the
observations and laboratory experiments are compared and discussed in
the context of recent astrochemical models. The conclusions of this
paper are given in Section \ref{sec.con}.

\section{Observations}
\label{sec.obs}

\subsection{The ALMA PILS survey}

We searched for methyl isocyanate in the Protostellar Interferometric
Line Survey (PILS) data, an unbiased spectral survey of the low-mass
protostellar binary IRAS16293 with ALMA. A full description and data
reduction of the survey is presented in
\citet{jorgensen2016}. Briefly, this survey covers a spectral range
from 329.147 to 362.896 GHz and was obtained with both the 12m array
and the Atacama Compact Array (ACA). The beam size ranges between
$\sim$0.4$\arcsec$ and 0.7$\arcsec$ depending on the configuration at
the time of the observations. The rms of the combined data sets is
about 7--10 mJy beam$^{-1}$ channel$^{-1}$, i.e., approximately 4--5 mJy
beam$^{-1}$ km s$^{-1}$. To facilitate the analysis, the combined data set used
in this paper was produced with a circular restoring beam of
0.5$\arcsec$ at a spectral resolution of 0.2 km s$^{-1}$.

Two positions are analysed in this study. The first position is offset
by one beam diameter ($\sim$0.5$\arcsec$) from the continuum peak of
source B in the south west direction ($\alpha_{\rm J2000}$=16$^{\rm h}$32$^{\rm m}$22$\fs$58, $\delta_{\rm J2000}$=-24$\degr$28$\arcmin$32.8$\arcsec$) \citep[see high resolution images in][]{baryshev2015}. Source B presents narrow
lines (FWHM $\sim$ 1 km\,s$^{-1}$). This position is found to be
optimal for line identifications, as the lines are particularly bright,
do not have strong absorption features toward the bright continuum of
source B, and do not suffer from high continuum optical depth
\citep{coutens2016,lykke2017,jorgensen2016}. In this paper, we also
analyse source A, which exhibits broader lines than source B making the
line identification quite challenging \citep{pineda2012}. The
linewidth varies, however, depending on the spatial separation from
this source. With an average FWHM of $\sim$ 2.5 km\,s$^{-1}$, the
position $\alpha_{\rm J2000}$=16$^{\rm h}$32$^{\rm m}$22$\fs$90,
$\delta_{\rm J2000}$=$-24 \degr 28 \arcmin 36.2 \arcsec$ appears to be
one of the best positions to search for new species toward source A
(0.3$''$ offset). At this position, the emission is centered at $v_{\rm
  LSR}$ of $\sim$0.8 km\,s$^{-1}$, blueshifted from the source A
velocity of $v_{\rm LSR}=3.2$ km s$^{-1}$
\citep{jorgensen2011}. 

\citet{bisschop2008} found HNCO and other
nitrogen containing species such as CH$_3$CN to be more prominent toward source A
than source B. Consequently, it is also interesting to check whether there exists a small
scale chemical differentiation among the N-bearing species for the two
sources.

\subsection{Results}

Methyl isocyanate is detected toward both components A and B of
IRAS16293. The identification is based on spectroscopic data from the
Cologne Database for Molecular Spectroscopy (CDMS,
\citealt{muller2001,muller2005}), taken from \citet{cernicharo2016} as
well as from \citet{koput1986}. CH$_3$NCO is an asymmetric rotor with
the NCO group lying at an angle of 140$^o$ and a low-lying ($\sim182$
cm$^{-1}$) C-N-C bending mode $\nu_b$. Moreover, the barrier to
internal rotation of the CH$_3$ group is low, only 21 cm$^{-1}$. The
pure rotational spectrum therefore has $A$ and $E$ torsional states
and vibrationally excited transitions can become detectable at temperatures
of a few hundred K. The labelling of the states used here refers to the quantum
numbers $J$ (rotational angular momentum), $K_a$ and $K_c$ (projection of angular momentum on the respective inertial axes) with internal rotation interactions
indicated by the quantum number $m$, with $m$=0 and $\pm$3 for the $A$
states and $m$=1, -2 and 4 for the $E$ states
\citep{halfen2015,cernicharo2016}.

\begin{figure*}
\begin{center}
\includegraphics[width=17.8cm]{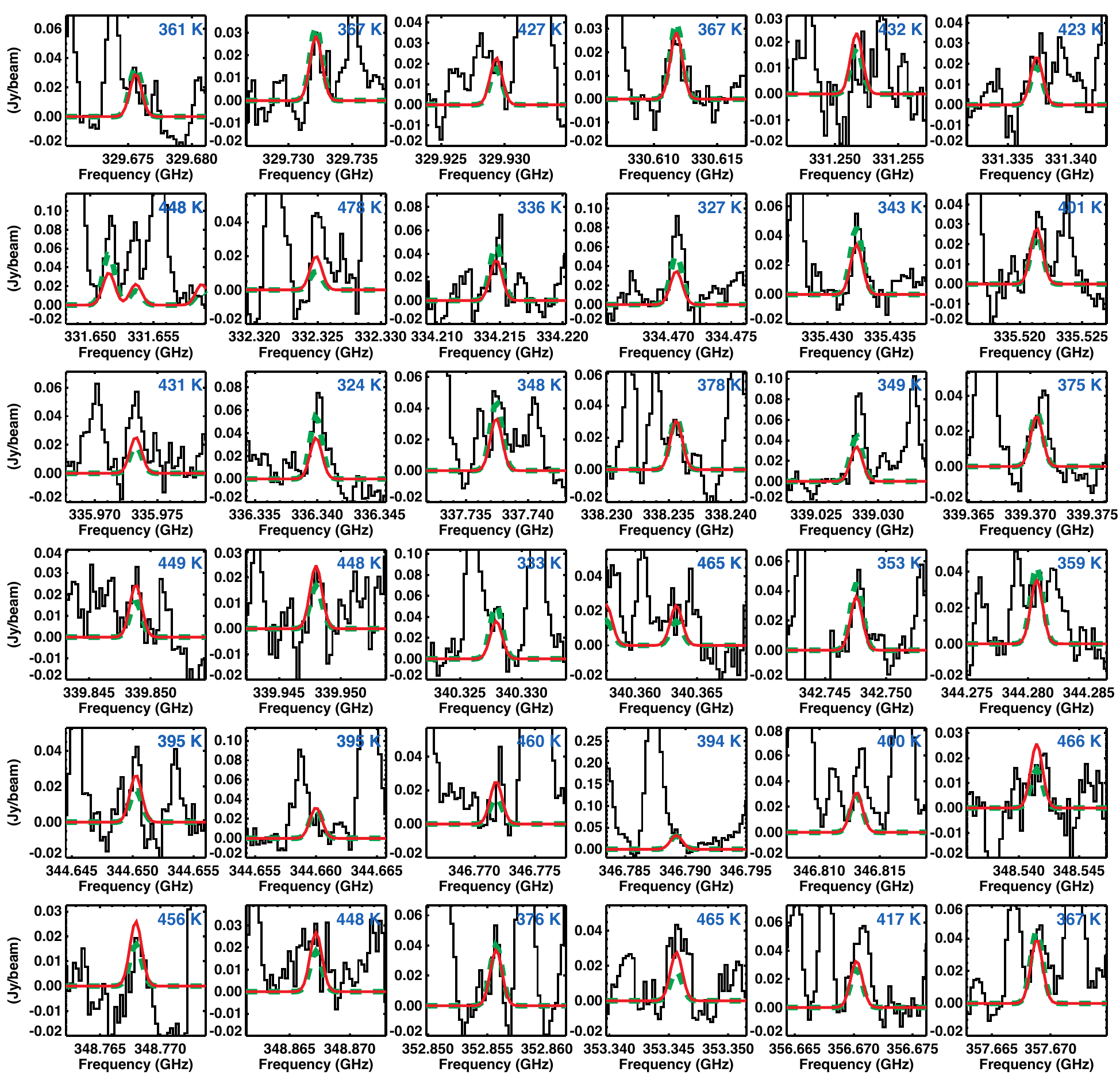}
\includegraphics[width=17.8cm]{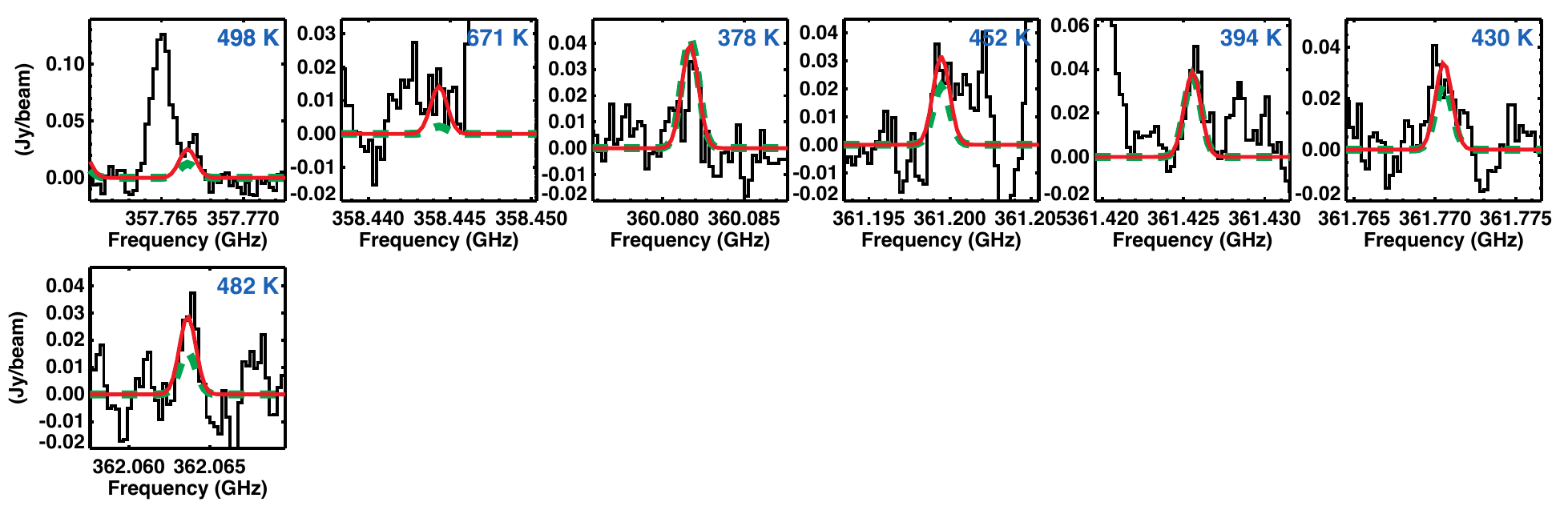}
\caption{Black: detected unblended lines of CH$_3$NCO toward source B. Red solid: best-fit model for $T_{\rm ex}$ = 300 K. Green dashed: best-fit model for $T_{\rm ex}$ = 100 K. The $E_{\rm up}$ values of the lines are indicated in blue in the right upper part of each panel.}
\label{fig_CH3NCO_vb0_B}
\end{center}
\end{figure*}

Using the CASSIS software\footnote{\url{http://cassis.irap.omp.eu/}},
we have been able to identify 43 unblended lines of CH$_3$NCO in the bending ground
state ($\nu_b$ = 0) with upper energy levels $E_{\rm up}$ ranging from
320 to 670 K toward source B (see Table \ref{table_obs}). Figure
\ref{fig_CH3NCO_vb0_B} shows the unblended lines detected toward this
component as well as the LTE modeling for two different excitation
temperatures at 100 and 300 K. Both excitation temperatures allow to reproduce the observations; only predicted transitions are observed. Our data are not sensitive to any
cold CH$_3$NCO component since lines with low $E_{\rm up}$ values are
missing in the spectral range of the PILS survey. Among the complex molecules that were detected and analysed towards source B, some (acetaldehyde, ethylene oxide) exhibit a relatively low excitation temperature of $\sim$125 K \citep{lykke2017}, while others (formamide, isocyanic acid, methanol, methyl formate, glycolaldehyde, ethylene glycol) show a higher excitation temperature of $\sim$ 300 K \citep{jorgensen2012,coutens2016,jorgensen2016}. Their spatial distribution is however rather similar and it is not possible to determine to which category CH3NCO belongs. The methyl isocyanate column density
is not very sensitive to the exact value of the excitation
temperature: assuming the same source size of 0.5$\arcsec$ as used in
previous PILS studies \citep{coutens2016,jorgensen2016,lykke2017}, the CH$_{3}$NCO
column density is found to be about 3\,$\times$\,10$^{15}$ cm$^{-2}$
and 4\,$\times$\,10$^{15}$ cm$^{-2}$ for $T_{\rm ex}$ = 300  and 100 K, respectively. All lines are optically thin. It is expected that for the same assumptions all column densities toward source B are accurate to better than 30\%. 

\begin{figure*}
\begin{center}
\includegraphics[width=18cm]{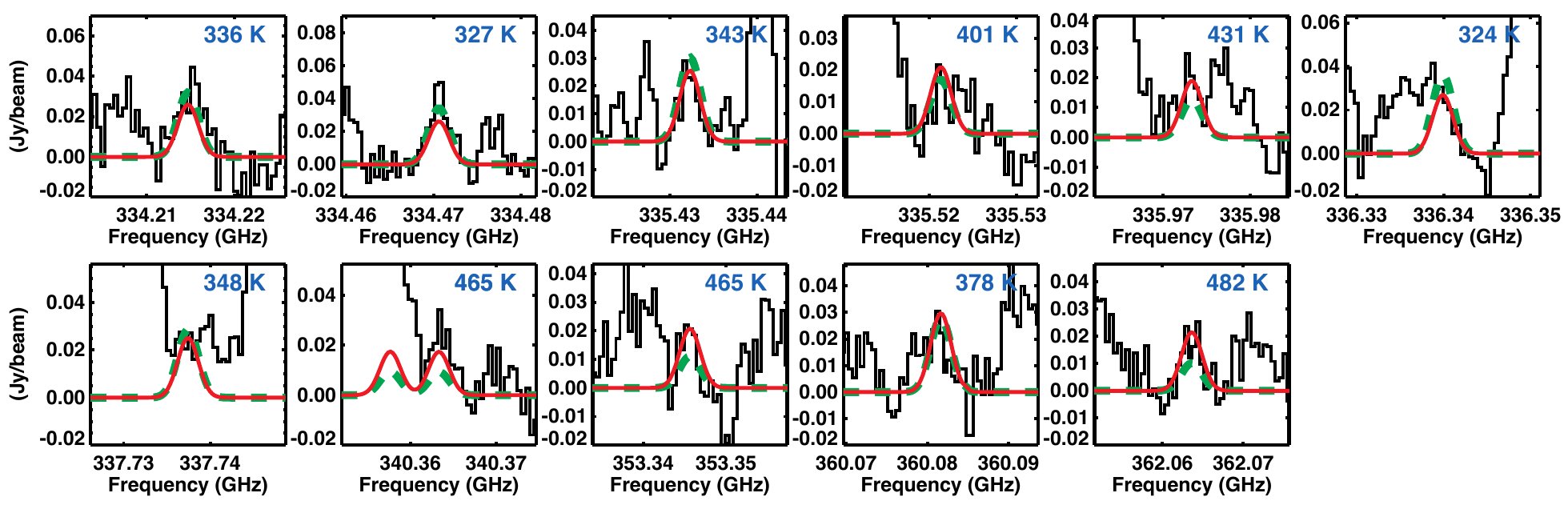}
\caption{Black: detected unblended lines of CH$_3$NCO toward source A. Red solid: best-fit model for $T_{\rm ex}$ = 300 K. Green dashed: best-fit model for $T_{\rm ex}$ = 100 K. The $E_{\rm up}$ values of the lines are indicated in blue in the right upper part of each panel.}
\label{fig_CH3NCO_vb0_A}
\end{center}
\end{figure*}

Toward source A, most of the lines are blended due to the larger
linewidths ($\sim$ 2.5 km\,s$^{-1}$). We can, however, identify 11
unblended lines of CH$_3$NCO (see Figure \ref{fig_CH3NCO_vb0_A}). A
column density of $\sim$ 6\,$\times$\,10$^{15}$ cm$^{-2}$ and $\sim$ 9\,$\times$\,10$^{15}$ cm$^{-2}$ (assuming a
source size of 0.5$\arcsec$) is in good agreement with
observations for excitation temperatures of 300 and 100 K, respectively, again with an
uncertainty of about 30\%. 

\begin{figure}
\begin{center}
\includegraphics[width=\hsize]{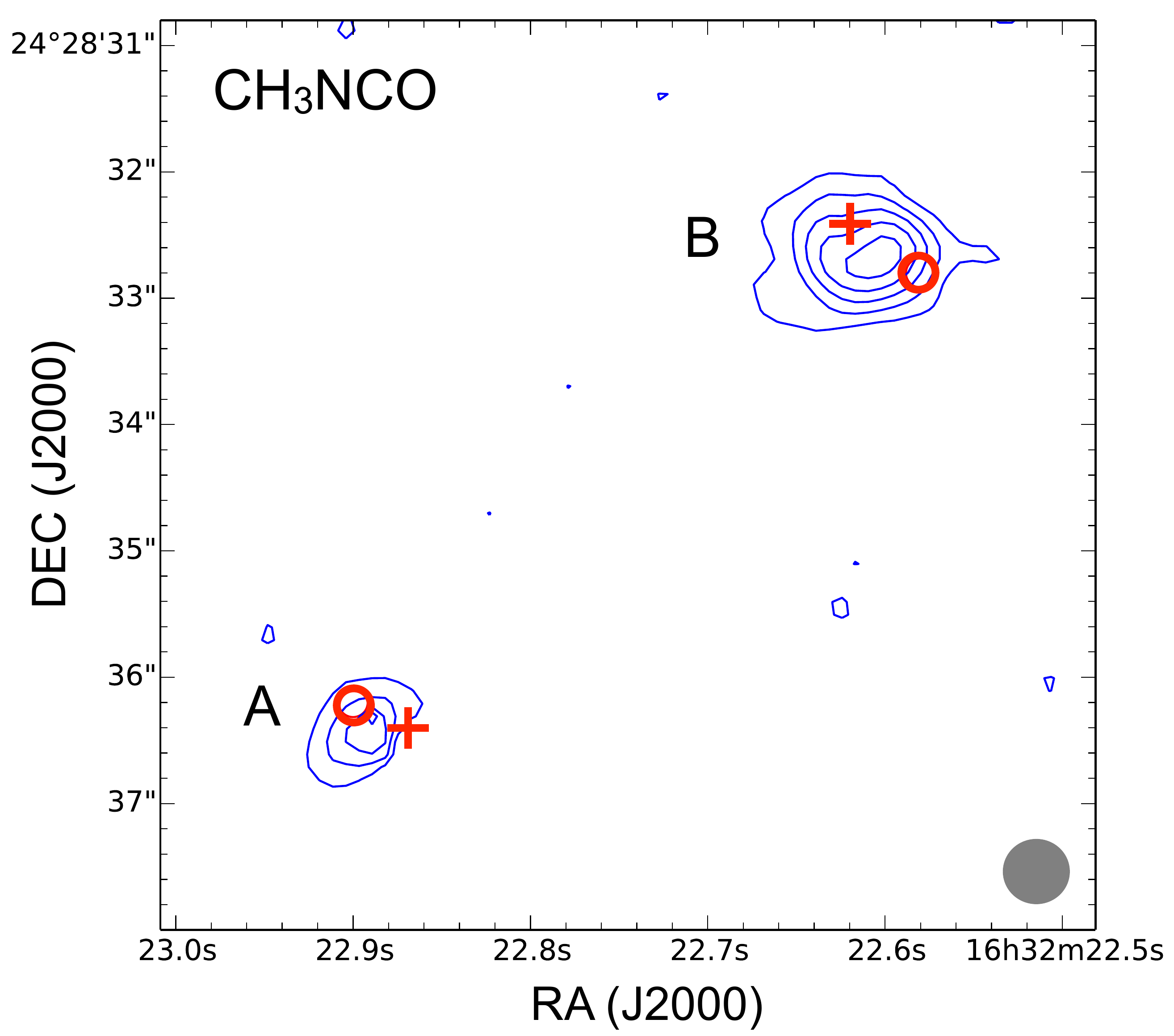}
\caption{Integrated intensity map of the CH$_{3}$NCO 39$_{0,39,0}$--38$_{0,38,0}$ transition at 336339.9 MHz and $E_{\rm up}$ = 323.7 K between 1.7 and 3.7 km\,s$^{-1}$. The positions of the continuum of source A (South East source) and source B (North West source) are indicated with red crosses, while the positions studied in this paper are indicated with red circles. 
The contour levels start at 5$\sigma$ with additional steps of 5$\sigma$. The circular restoring beam of 0.5$\arcsec$ size is indicated in grey in the right hand lower corner.}
\label{fig_map_MIC}
\end{center}
\end{figure}

At high excitation temperatures, rotational levels in the first excited
bending state ($\nu_b$=1) may be populated as well and predictions for possible transitions are shown
in Figure \ref{fig.ch3nco_vb1}. For $T_{\rm ex}$ = 300 K, some faint
lines can indeed be tentatively attributed to CH$_{3}$NCO $\nu_b$=1 transitions
toward source B. An integrated intensity map of one of the
brightest CH$_{3}$NCO lines, the 39$_{0,39,0}$--38$_{0,38,0}$ transition at 336339.9 MHz, is presented in Figure
\ref{fig_map_MIC}. Similarly to other complex molecules, the emission
is quite compact with a size of $\sim$60 AU radius and centred near the two sources, with little difference between them
\citep{coutens2016,jorgensen2016,lykke2017}. For source B, the emission
is somewhat offset due to the continuum becoming optically thick on
source.

\begin{table*}
\centering
\caption{Comparison of molecular abundance ratios toward different sources}
\label{tab.comp_obs}
\begin{tabular}{lccccc}
\hline \hline
Source  & HNCO/CH$_3$NCO & CH$_3$CN/CH$_3$NCO & CH$_3$OH/CH$_3$NCO & Reference \\
\hline
IRAS16293 B & 12 & 8 & 3333 & \multirow{2}{*}{This work}\\
IRAS16293 A & 4 & 9 & 3200 & \\
\hline
Orion KL A & 15 & 10 & 400 & \multirow{3}{*}{\citet{cernicharo2016}}\\
Orion KL B & 15 & 25 & 225 & \\
Sgr B2(N1) & 40 & 50 & 40 & \\
\hline
Sgr B2(N2) &  9	& 10 & 182 & \citet{belloche2016,belloche2017} \\
\hline
\multirow{2}{*}{Comet 67P/C-G} & \multirow{2}{*}{>0.2} & \multirow{2}{*}{>0.2} & \multirow{2}{*}{--} & \citet{goesmann2015} \\
 & & & & Altwegg et al. in prep. \\
\hline
\end{tabular}
\\
\end{table*}

We also searched for spectral evidence of two isomers of methyl isocyanate - acetonitrile oxide (CH$_3$CNO, \citealt{winnewisser1982}) and methyl cyanate (CH$_3$OCN, \citealt{sakaizumi1990,kolesnikova2016}) - but the corresponding transitions were not found in the PILS data set. From this non-detection 
$3\sigma$ upper limits of 5\,$\times$\,10$^{13}$
cm$^{-2}$ and 5\,$\times$\,10$^{14}$ cm$^{-2}$ are derived, respectively (assuming an excitation temperature of 100 K). These isomers are consequently less abundant than methyl
isocyanate by at least two and one orders of magnitude, in agreement with recent
findings toward Orion KL \citep{cernicharo2016,kolesnikova2016}.

In view of the important role that HNCO and HOCN may play in the
formation of CH$_{3}$NCO, we also searched for spectral signatures of these precursor species. The
analysis of the PILS data for HNCO toward source B is presented in
\citet{coutens2016}. The HNCO lines are optically thick, so lines of
the isotopologue HN$^{13}$CO were analysed and the HNCO column density
was derived assuming a $^{12}$C/$^{13}$C ratio of 68
\citep{milam2005}.
To get precise abundance ratios, we re-analysed the HN$^{13}$CO data using the same dataset (with the circular restoring beam of 0.5$\arcsec$) and obtained a HNCO column density of 3.7\,$\times$\,10$^{16}$ cm$^{-2}$, which is within the 30\% uncertainty range. The resulting HNCO/CH$_3$NCO abundance ratio is about 12
toward source B with an uncertainty of less than a factor 2.
Within the error margins, this is similar to the value derived in Orion KL
\citep{cernicharo2016} or Sgr B2(N2) \citep{belloche2016,belloche2017}. 
For source A, the column density of HNCO is estimated to be
about 3.4\,$\times$\,10$^{16}$ cm$^{-2}$ ($T_{\rm ex}$ = 100 K). 
The corresponding HNCO/CH$_3$NCO ratio is consequently about 4 toward this
component, with a somewhat larger uncertainty of a factor of 3 due to the difficulty of deriving a precise column density for HNCO because of line blending. Overall, we can
conclude that the two HNCO/CH$_3$NCO ratios are comparable toward the two components of the binary.

Whereas HNCO is readily identified, this is not the case for HOCN.  No
HOCN lines were detected and a $3\sigma$ upper limit of
3\,$\times$\,10$^{13}$ cm$^{-2}$ is derived toward source B. HOCN is
consequently less abundant than HNCO and CH$_{3}$NCO by at least 3 and 2 orders
of magnitude, respectively. Therefore, it is highly unlikely that the
gas-phase formation pathway involving HOCN, as proposed by
\citet{halfen2015} in Eq.\ \ref{eq.halfen1} and \ref{eq.halfen2}, contributes significantly to the
production of methyl isocyanate in this source.

We have also determined the abundance ratios of CH$_{3}$NCO with
respect to CH$_3$OH and CH$_3$CN for comparison with other sources
in which methyl isocyanate has been detected (see Table
\ref{tab.comp_obs}). The column density of CH$_3$OH was estimated based on CH$_3^{18}$OH by J\o rgensen et al. (in prep.) for the one beam offset position toward source B ($\sim$ 1\,$\times$\,10$^{19}$ cm$^{-2}$) using a $^{16}$O/$^{18}$O ratio of 560 \citep{wilson1994}. With the same assumptions, we estimate a column density of CH$_3$OH of $\sim$ 2\,$\times$10$^{19}$ cm$^{-2}$ in source A. CH$_3$CN was analysed by Calcutt et al. (in prep.) toward both source A ($\sim$
8\,$\times$\,10$^{16}$ cm$^{-2}$) and source B ($\sim$
3\,$\times$\,10$^{16}$ cm$^{-2}$). Similarly to the HNCO/CH$_3$NCO ratio, the
abundance ratio of CH$_3$CN/CH$_3$NCO is comparable to the values
found in Orion KL \citep{cernicharo2016} and lower than
toward Sgr B2(N1), but again comparable to Sgr B2(N2) \citep{belloche2016,belloche2017}. Methanol is, however, clearly more abundant than methyl isocyanate toward IRAS16293 than toward Orion KL and Sgr B2.

The HNCO/CH$_3$NCO and CH$_3$CN/CH$_3$NCO abundance ratios derived in
IRAS16293 are much larger than the lower limits found in comet 67P
($>$0.2). A proper comparison awaits the new results derived with the
ROSINA instrument (Altwegg et al. in prep.).

\section{Laboratory experiments}
\label{sec.lab}

Grain surface formation routes of complex molecules have been investigated
experimentally for many years using cryogenic set-ups to mimic specific chemical processes under fully controlled laboratory conditions \citep[see][for
  reviews]{herbstdishoeck2009,linnartz2015}. In the present work the
formation of methyl isocyanate is investigated by energetically processing solid-state
CH$_{4}$:HNCO mixtures with vacuum-UV radiation. VUV irradiated methane ice is known to efficiently yield methyl
radicals \citep{bossa2015}, and these radicals are expected to further react through surface diffusion with HNCO
to form CH$_{3}$NCO, as proposed by \citet{goesmann2015}
and \citet{cernicharo2016}, reaction \ref{eq.cernicharo1}. The Cryogenic Photoproduct Analysis Device
2 (CryoPAD2) in the Sackler Laboratory for Astrophysics is used to perform the measurements to investigate the role of methylation of HNCO in methyl isocyanate formation. A short description of this set-up, experimental procedure and
reference data is given below.

Note that in these experiments CH$_4$ is used as the parent of CH$_3$
but in interstellar space methyl radicals active in the ice may also originate from CH$_{3}$OH dissociation \citep{oberg2009a} or from direct CH$_{3}$ accretion from the gas-phase. The main aim of this section is to
investigate whether or not the proposed solid-state reaction 
as shown in reaction \ref{eq.cernicharo1} can proceed at temperatures typical for dark cloud conditions in the ISM, i.e., independent of the exact precursor species. 

\subsection{Set-up and method}

CryoPAD2 consists of a central chamber operating under ultra-high
vacuum conditions ($P \simeq$ 10$^{-10}$ mbar), to which various
instruments are attached. On the top a cryostat is mounted which cools a
gold-coated reflective surface down to 12 K. The temperature of this
surface can be controlled through resistive heating and varied
between 12 and 300 K, with an absolute temperature accuracy better than 1 K. Pure and mixed gases are prepared separately in a
gas-mixing system which is attached to a high-precision leak valve to
the main chamber and which doses the gases onto the cooled surface. The
deposited ice samples are irradiated with VUV radiation from a Microwave Discharge
Hydrogen-flow Lamp (MDHL), which emits radiation peaking at 121 nm and
a region between 140 to 160 nm, corresponding to photon energies in
the range of 7.5 to 10.2 eV \citep{chen2014,ligterink2015}. The total lamp flux
is (1.1$\pm$0.1)\,$\times$\,10$^{14}$ photons s$^{-1}$ that is determined using a calibrated NIST diode. CryoPAD2 possesses two diagnostic tools to monitor the VUV induced processes in the ice. The beam of a
Fourier-Transform IR Spectrometer (FTIRS) is directed under grazing
incidence onto the gold-coated surface, in order to perform Reflection
Absorption IR Spectroscopy (RAIRS). The sample preparation and chemical
changes under influence of VUV radiation are monitored by recording vibrational fingerprint spectra of molecules on the surface. To decrease the pertubing role of atmospheric CO$_{2}$ and H$_{2}$O, the system is purged with filtered dry air. The second instrument is
a highly sensitive Quadrupole Mass Spectrometer (QMS), with an ionization element at 70~eV, which is able to trace gas-phase
molecules in the chamber that are released from the ice surface upon linear heating during a Temperature Programmed Desorption (TPD) experiment. TPD is a very useful method that allows to identify desorbing species through their specific desorption temperature and mass fragmentation pattern. Unambiguous identifications become possible through the use of isotopologues and searching for the corresponding mass shifts of specific fragments. Obviously, TPD comes with the destruction of the ice.

In the experiments CH$_{4}$ (Linde Gas, 99.995\% purity),
$^{13}$CH$_{4}$ (Sigma-Aldrich, 99\% purity) and HNCO are used. HNCO
is produced from cyuranic acid (Sigma-Aldrich, 98\% purity), the solid
trimer of HNCO, following the protocol described in
\citet{vanbroekhuizen2004}. Impurities of the HNCO production process are
removed by freeze-thaw cycles to obtain a HNCO purity of >99\%.

For the experiments samples of pure HNCO and methane, and mixtures of
$^{13}$CH$_{4}$/CH$_{4}$:HNCO at 5:1 ratio are prepared. This ratio
is within a factor of 2 of that observed for interstellar ices
\citep{oberg2011} but is particularly chosen to create a large abundance of CH$_3$
radicals to test whether reaction \ref{eq.cernicharo1} proceeds or not. Homogeneously mixed ices are grown on the surface at 20 K and irradiated with a total fluence
of $\sim$3.3$\times$10$^{17}$ photons. During irradiation of the
sample, IR spectra are continuously recorded at 1 cm$^{-1}$
resolution. After the irradiation TPD is started, while still
recording IR spectra.

The strongest vibrational features of solid methyl isocyanate are
found between 2320 and 2280 cm$^{-1}$ ($\sim$4.34 $\mu$m) for the
-N=C=O antisymmetric stretching vibration and overtone 2$\nu_{7}$ CH$_{3}$
rocking mode. \citet{sullivan1994} lists these at 2320, 2280, 2270, 2240 and 2228 cm$^{-1}$, with 2280 cm$^{-1}$ being the strongest band. \citet{zhou2009} positioned all bands around 2300 cm$^{-1}$ and \citet{reva2010} puts the band for methyl isocyanate in a nitrogen matrix at 2334.7, 2307.9, 2288.9, 2265.2 and 2259.7 cm$^{-1}$, finding the strongest transition at 2288.9 cm$^{-1}$. The region around these bands is used to monitor CH$_{3}$NCO formation in the ice. We focus on the region between 2400 -- 2100 cm$^{-1}$. From previous experiments it is known that CO$_{2}$ (2340 cm$^{-1}$),
OCN$^{-}$ (2165 cm$^{-1}$) and CO (2135 cm$^{-1}$) are produced from
HNCO (2265 cm$^{-1}$) upon irradiation and these photo-products also have spectral features in the region characterizing methyl isocyanate
\citep{raunier2004,vanbroekhuizen2004}. Other known products of HNCO
irradiation are formamide, urea and formaldehyde, which do not have any interfering IR
features in the region of interest. Energetic processing of methane
does not yield products that show up in the region of
interest. Products that are formed from methane are mainly ethane and
to a lesser extent ethene and ethyn
\citep{bennett2006,bossa2015}. These species are seen in our
spectra at their appropriate frequencies. The abundant formation of ethane demonstrates that CH$_3$ is
indeed produced in the experiments, since ethane is the direct product of methyl-radical recombination.

In order to identify methyl isocyanate in the gas-phase using TPD, the mass
fragmentation pattern available from the NIST database\footnote{NIST
  Mass Spec Data Center, S.E. Stein, director, "Mass Spectra" in NIST
  Chemistry WebBook, NIST Standard Reference Database Number 69,
  Eds. P.J. Lindstrom and W.G. Mallard, National Institute of
  Standards and Technology, Gaithersburg MD, 20899,
  http://webbook.nist.gov.} is used. The fragmentation pattern at 70~eV comprises unique peaks at $m/z$ = 57 and 56  (hereafter also called the
primary and secondary mass peak), which have a $m/z$ 57:56 ratio of
5:2 and these will be used as main TPD mass tracers. Known products of pure HNCO and methane irradiation does not have a
mass fragmentation peak at $m/z$ = 57 (see also Appendix \ref{sec.mass_CH4_HNCO}).

It is important to mention that methyl isocyanate is severely toxic and specialised laboratories and equipment are needed to work with this substance. This complicates the extensive use of this species and for this reason additional experiments, starting from the pure precursor, have not been performed.

\subsection{Results -- IR Spectra}

\begin{figure}
\begin{center}
\includegraphics[width=\hsize]{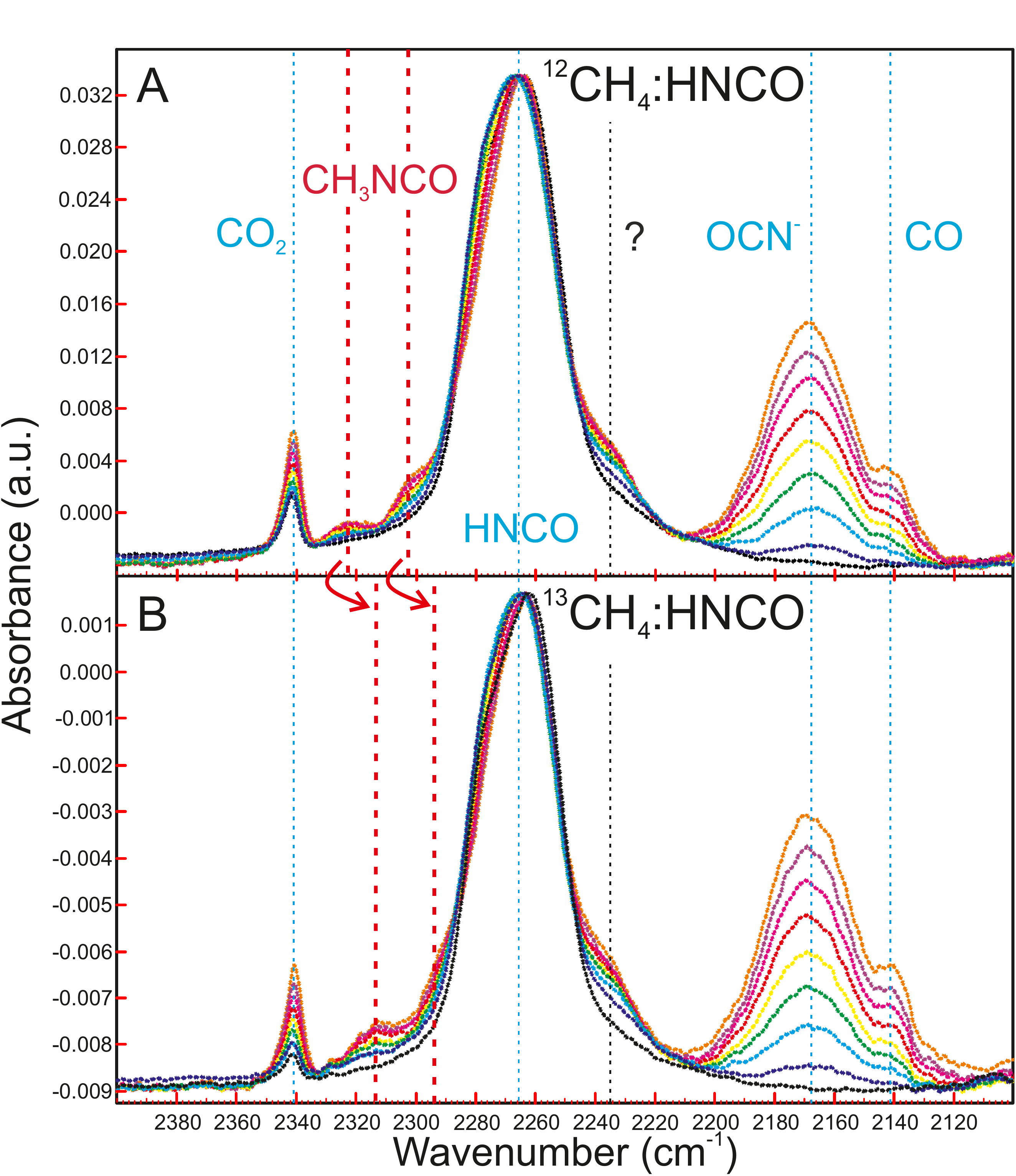}
\caption{IR spectra taken over time for the $^{12}$CH$_{4}$:HNCO (A) and $^{13}$CH$_{4}$:HNCO (B) mixture. HNCO and the products CO$_{2}$, OCN$^{-}$ and CO are listed (blue). Spectroscopic features that coincide with CH$_{3}$NCO are found at the red lines and show a clear shift with the $^{13}$CH$_{4}$ isotopologue. One unidentified peak is found in the right wing of the HNCO peak, indicated by a question mark. }
\label{fig.FTIR_plot}
\end{center}
\end{figure}

Figure \ref{fig.FTIR_plot} presents the IR spectra recorded during the first 1$\times$10$^{17}$ photons
irradiating the $^{12/13}$CH$_{4}$:HNCO samples. All
spectra are normalized to the HNCO peak. Three known spectroscopic
features of CO$_{2}$, OCN$^{-}$ and CO (blue) show up during
irradiation. Also, two new features become visible around 2300 cm$^{-1}$
(red), which do not show up while processing samples of pure HNCO or
CH$_{4}$. Also a clear red shift of about 10 cm$^{-1}$ of the two features is seen between the
sample of $^{12}$CH$_{4}$ and $^{13}$CH$_{4}$, moving transitions at 2322 and
2303 cm$^{-1}$ to 2313 and 2294 cm$^{-1}$. These spectroscopic features
are therefore the result of a product formed in the reaction between methane and
isocyanic acid, and, since they are found close to known
CH$_{3}$NCO features (given by \citet{sullivan1994} and \citet{reva2010}), are plausibly identified with methyl isocyanate. Another feature is seen in the wing of the
HNCO peak around 2235 cm$^{-1}$, which does not clearly shift with
methane isotopologues. The origin of this band is unclear. 

Bandstrength values for methyl isocyanate are not available from the literature, however, a rough indication of the amount of formed methyl isocyanate versus lost HNCO can be given by making the assumption that the bandstrength of the NCO antisymmetric stretch vibration of methyl isocyanate equals that of the corresponding vibration of HNCO. To obtain the ratio, the integrated area of the 2303 cm$^{-1}$ feature is divided by the integrated loss area of the 2265 cm$^{-1}$ HNCO band for a number of spectra. A ratio of $N$(HNCO)/$N$(CH$_{3}$NCO) = 100 -- 200 is found, which is about an order of magnitude higher than the ratio observed toward IRAS16293. It should be noted, that it is not $a priori$ clear whether solid state laboratory and gas phase astronomical abundances can be directly compared (see e.g. Chuang et al. 2017). It is likely that ongoing gas phase reactions, for one or the other species, change the solid state to gas phase ratios. It is also possible that non-linear RAIRS effects can offset the column density \citep{teolis2007} or that the CH$_{3}$NCO bandstrength is significantly different. In a same manner CH$_{3}$NCO  photodestruction  may affect the overall abundances.

\subsection{Results -- temperature programmed desorption}

\begin{figure}
\begin{center}
\includegraphics[width=\hsize]{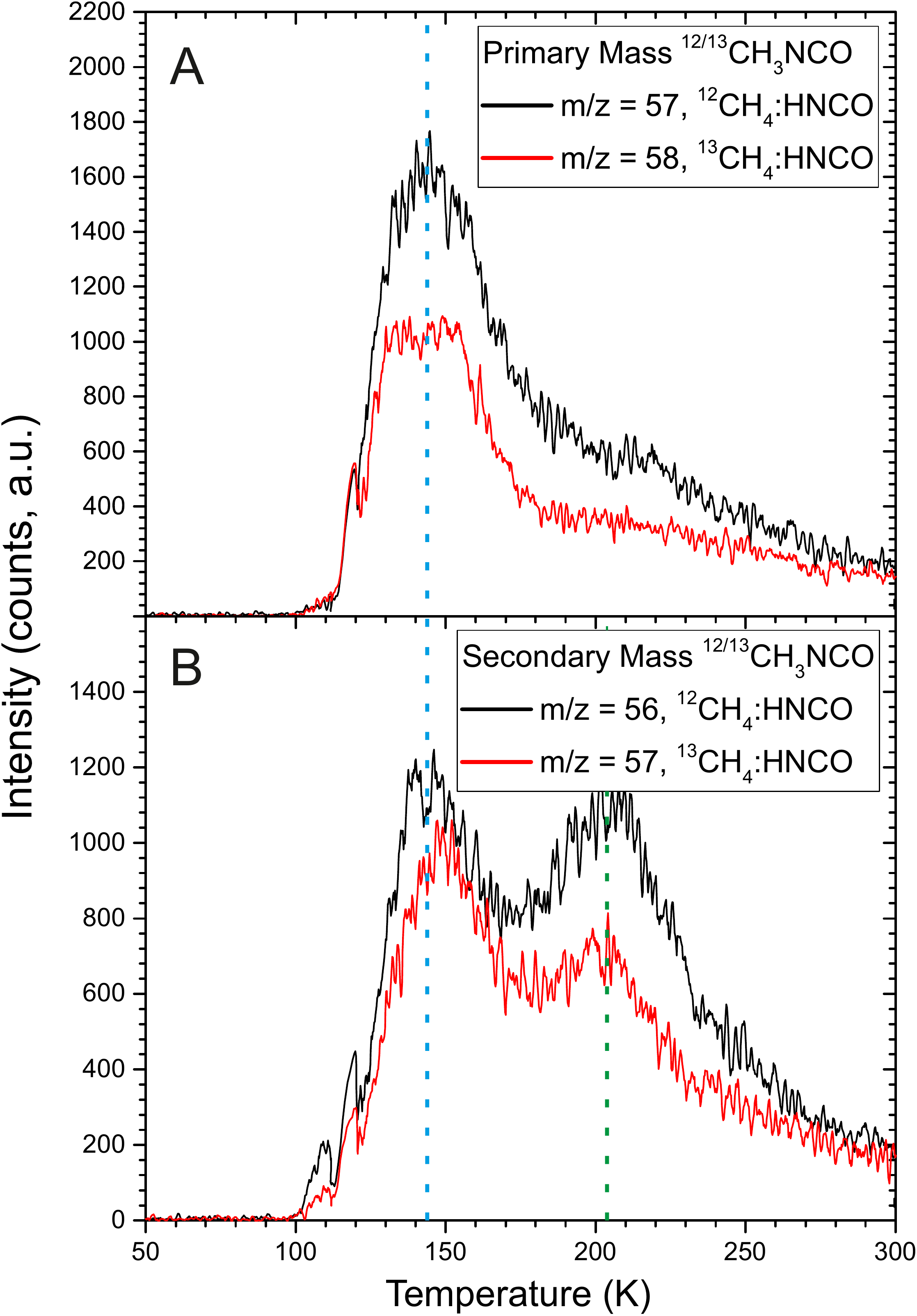}
\caption{TPD trace of the primary (A) and secondary (B) masses of $^{12/13}$CH$_{3}$NCO. Both the primary and secondary mass are seen being released around 145 K (blue line). The secondary mass shows another release peak around 205 K (green), which is suspected to also contribute to first release peak, thus altering respectively the $m/z$ 57/56 and 58/57 ratio.}
\label{fig.mass_trace}
\end{center}
\end{figure}

The desorption temperature of methyl isocyanate has not been reported in the
literature, but the TPD traces of our experiments on UV processed CH$_{4}$:HNCO ices do show the combined
release of the primary and secondary masses of methyl isocyanate
$m$=57 and 56 at 145 K (Figure \ref{fig.mass_trace}). The verification experiment of pure HNCO did not show these masses and only $m/z$ = 56 (and no 57) was seen after the irradiation of pure methane ice, releasing at 105 K (Figure \ref{fig.mass_CH4_HNCO} in the Appendix). At the 145 K desorption peak the primary over secondary
mass ratio is around 1-1.5, lower than the value of 2.5 suggested by
NIST upon 70 eV electron impact ionization. The NIST calibration values are a good indicator of the values to be expected, but are to some extent set-up specific. In such cases, the expected reactant can be deposited directly and the fragmentation pattern can be studied adapted to the set-up in use. This is unfortunately not possible here due to the aforementioned toxicity of methyl isocyanate. Moreover, contributions from other reaction products cannot be fully excluded. In
Figure \ref{fig.mass_trace}B a second desorption peak is found around
205 K, which is seen only as a shoulder in panel A. It is unlikely that this peak is associated with methyl isocyanate, instead it shows that the chemical network involves the formation of other species as well. In fact, since the additional product(s) have a mass fragmentation
pattern which contributes to the secondary mass of methyl isocyanate,
this could explain why the primary/secondary mass ratio does not
exactly match with that given by NIST. Additionally, the unidentified product of pure methane irradiation could be trapped in the ice and contribute to other desorption peaks. Therefore the 145 K desorption peak is still consistent with methyl isocyanate. There is also no
other candidate molecule with a primary mass of 57 in the NIST
database that could plausibly explain the TPD spectra. 

CryoPAD2 offers the unique feature to combine IR and TPD data, i.e., at the moment a specific ice feature starts thermally desorbing, the IR signal starts decreasing and simultaneously the mass signal is increasing. In the measurements presented here, this effect is not as clearly visible as in previous studies (e.g. \citet{oberg2009a}), but this may be partly due to the low final S/N of the feature, making it difficult to link RAIRS and TPD signals. The present data are as accurate as possible, but obviously their interpretation would benefit from focused experiments determining the thermal desorption peak of pure CH$_{3}$NCO and the IR bandstrengths.

\section{Linking observations and laboratory data}
\label{sec.dis}

Methyl isocyanate is detected in a significant abundance toward both components of
IRAS16293 with an excitation temperature of at least 100 K. The two
isomers, CH$_{3}$CNO and CH$_{3}$OCN, are not detected. Abundance
ratios with respect to HNCO and CH$_{3}$CN are found to be similar to
those found for Orion KL and comparable within a factor of a few to
Sgr B2, making a common formation pathway in these three sources
likely. Moreover, no chemical differentiation between source A and B
is found. Gas-phase production of CH$_{3}$NCO via HOCN (reaction
\ref{eq.halfen1}) can be seen as insignificant due to the low upper
limit on this molecule, but a gas-phase route involving HNCO cannot be excluded. 

The laboratory experiments on the formation of CH$_{3}$NCO
strongly suggest that a solid-state formation scheme is possible, starting from two astronomically relevant precursor species, CH$_{4}$ and HNCO. The proposed reaction
CH$_{3}$ + HNCO $\rightarrow$ CH$_{3}$NCO + H indeed seems to proceed, confirming a solid state reaction pathway, but other routes such as hydrogen stripping CH$_{3}$ + HNCO
$\rightarrow$ CH$_{4}$ + NCO may take place in parallel. Reactions
involving the OCN$^{-}$ anion, which is abundantly formed in
these experiments, provide alternative routes. Irradiation of a sample of
OCN$^{-}$:CH$_{4}$, where OCN$^{-}$ is formed via the acid-base
reaction of HNCO with NH$_{3}$, will be interesting to study in future
work since this anion is also a well known component of interstellar ice \citep{boogert2015}.

\citet{belloche2017} have incorporated reactions \ref{eq.belloche1} and \ref{eq.belloche2} (see Section \ref{sec.int}) into the large
gas-grain model MAGICKAL \citep{garrod2013} and simulated the cold
collapse + warm-up phase of a hot core region. Although tailored to
the high-mass source SgrB2(N) many of the model features are also
expected to be valid for the case of IRAS16293. In this model,
HNCO is formed by reaction of NH + CO, a route that has been
demonstrated experimentally by \citet{fedoseev2015,fedoseev2016}. Depending on assumptions about the barriers of the grain surface reactions involved in the formation of
HNCO and CH$_3$NCO, abundance ratios HNCO/CH$_3$NCO of 3-50 following
ice sublimation are readily found in the models, consistent with the
IRAS16293 observations.

\section{Conclusions}
\label{sec.con}
The main observational and experimental conclusions are listed below:

\begin{enumerate}
\item Methyl isocyanate is detected for the first time toward a low mass protostar, IRAS16293, on solar system scales (emission radius of 60 AU). Column densities of $\sim$ (3--4)\,$\times$\,10$^{15}$ and $\sim$ (6--9)\,$\times$\,10$^{15}$ cm$^{-2}$ are obtained toward source B and A, respectively, yielding $N$(HNCO)/$N$(CH$_{3}$NCO) = 12 and 4, with no significant variation between the two sources. 
\item The abundance ratios of CH$_3$NCO relative to the N-bearing species HNCO and CH$_3$CN are similar to those found toward Orion KL and deviate by at most an order of magnitude from Sgr B2.
\item The isomers of methyl isocyanate, CH$_{3}$CNO and CH$_{3}$OCN, are not detected. These species are less abundant than CH$_3$NCO by at least a factor 100 and 10, respectively.
\item HOCN is not found down to <3 $\times$ 10$^{13}$ cm$^{-2}$, giving $N$(HOCN)/$N$(HNCO) > 1000, which makes this an insignificant gas-phase precursor to methyl isocyanate in IRAS16293.
\item Laboratory experiments strongly suggest that it is possible to form CH$_{3}$NCO on an icy surface, irradiating CH$_{4}$ and HNCO as astronomically relevant precursor species with VUV light, generating methyl radicals as reactive intermediates to form methyl isocyanate.
\end{enumerate}  

The detection of CH$_{3}$NCO adds to the growing list
of complex molecules known to be present around solar mass protostars,
showing that the ingredients for prebiotic molecules are
abundant. Future deeper searches for even more complex molecules
relevant for the origin of life are warranted.

\section*{Acknowledgements}

This paper makes use of the following ALMA data:
ADS/JAO.ALMA\#2013.1.00278.S. ALMA is a partnership of ESO
(representing its member states), NSF (USA) and NINS (Japan), together
with NRC (Canada) and NSC and ASIAA (Taiwan), in cooperation with the
Republic of Chile. The Joint ALMA Observatory is operated by ESO,
AUI/NRAO and NAOJ. Astrochemistry in Leiden is supported by the European Union A-ERC
grant 291141 CHEMPLAN, by the Netherlands Research School for
Astronomy (NOVA) and by a Royal Netherlands Academy of Arts and Sciences
(KNAW) professor prize. CryoPAD2 was realized with NOVA and NWO (Netherlands Organisation for Scientific Research) grants. The work of A.C. was funded by the STFC grant ST/M001334/1. The group of J.K.J. acknowledges support from a Lundbeck Foundation Group Leader Fellowship, as well as the ERC under the European Union's Horizon 2020 research and innovation programme through ERC Consolidator Grant S4F
(grant agreement No 646908). Research at the Centre for Star and Planet Formation is funded by the Danish National Research Foundation.
The authors of this paper thank A. Belloche (Max-Planck-Institut f\"ur Radioastronomie) for useful input to the paper. We would like to thank the referee for the constructive comments.




\bibliographystyle{mnras}
\bibliography{lib} 

\begin{thebibliography}{}
\makeatletter
\relax
\def\mn@urlcharsother{\let\do\@makeother \do\$\do\&\do\#\do\^\do\_\do\%\do\~}
\def\mn@doi{\begingroup\mn@urlcharsother \@ifnextchar [ {\mn@doi@}
  {\mn@doi@[]}}
\def\mn@doi@[#1]#2{\def\@tempa{#1}\ifx\@tempa\@empty \href
  {http://dx.doi.org/#2} {doi:#2}\else \href {http://dx.doi.org/#2} {#1}\fi
  \endgroup}
\def\mn@eprint#1#2{\mn@eprint@#1:#2::\@nil}
\def\mn@eprint@arXiv#1{\href {http://arxiv.org/abs/#1} {{\tt arXiv:#1}}}
\def\mn@eprint@dblp#1{\href {http://dblp.uni-trier.de/rec/bibtex/#1.xml}
  {dblp:#1}}
\def\mn@eprint@#1:#2:#3:#4\@nil{\def\@tempa {#1}\def\@tempb {#2}\def\@tempc
  {#3}\ifx \@tempc \@empty \let \@tempc \@tempb \let \@tempb \@tempa \fi \ifx
  \@tempb \@empty \def\@tempb {arXiv}\fi \@ifundefined
  {mn@eprint@\@tempb}{\@tempb:\@tempc}{\expandafter \expandafter \csname
  mn@eprint@\@tempb\endcsname \expandafter{\@tempc}}}

\bibitem[\protect\citeauthoryear{{Baryshev} et~al.,}{{Baryshev}
  et~al.}{2015}]{baryshev2015}
{Baryshev} A.~M.,  et~al., 2015, \mn@doi [\aap] {10.1051/0004-6361/201425529},
  \href {http://adsabs.harvard.edu/abs/2015A%26A...577A.129B} {577, A129}

\bibitem[\protect\citeauthoryear{{Belloche}, {M{\"u}ller}, {Menten}, {Schilke}
  \& {Comito}}{{Belloche} et~al.}{2013}]{belloche2013}
{Belloche} A.,  {M{\"u}ller} H.~S.~P.,  {Menten} K.~M.,  {Schilke} P.,
  {Comito} C.,  2013, \mn@doi [\aap] {10.1051/0004-6361/201321096}, \href
  {http://adsabs.harvard.edu/abs/2013A%26A...559A..47B} {559, A47}

\bibitem[\protect\citeauthoryear{{Belloche}, {Garrod}, {M{\"u}ller}  \&
  {Menten}}{{Belloche} et~al.}{2014}]{belloche2014}
{Belloche} A.,  {Garrod} R.~T.,  {M{\"u}ller} H.~S.~P.,   {Menten} K.~M.,
  2014, \mn@doi [Science] {10.1126/science.1256678}, \href
  {http://adsabs.harvard.edu/abs/2014Sci...345.1584B} {345, 1584}

\bibitem[\protect\citeauthoryear{{Belloche}, {M{\"u}ller}, {Garrod}  \&
  {Menten}}{{Belloche} et~al.}{2016}]{belloche2016}
{Belloche} A.,  {M{\"u}ller} H.~S.~P.,  {Garrod} R.~T.,   {Menten} K.~M.,
  2016, \mn@doi [\aap] {10.1051/0004-6361/201527268}, \href
  {http://adsabs.harvard.edu/abs/2016A%26A...587A..91B} {587, A91}

\bibitem[\protect\citeauthoryear{{Belloche} et~al.,}{{Belloche}
  et~al.}{2017}]{belloche2017}
{Belloche} A.,  et~al., 2017, accepted in A\&A, \href
  {http://adsabs.harvard.edu/abs/2017arXiv170104640B} {}

\bibitem[\protect\citeauthoryear{{Bennett}, {Jamieson}, {Osamura}  \&
  {Kaiser}}{{Bennett} et~al.}{2006}]{bennett2006}
{Bennett} C.~J.,  {Jamieson} C.~S.,  {Osamura} Y.,   {Kaiser} R.~I.,  2006,
  \mn@doi [\apj] {10.1086/508561}, \href
  {http://adsabs.harvard.edu/abs/2006ApJ...653..792B} {653, 792}

\bibitem[\protect\citeauthoryear{{Bisschop}, {J{\o}rgensen}, {van Dishoeck}  \&
  {de Wachter}}{{Bisschop} et~al.}{2007}]{bisschop2007}
{Bisschop} S.~E.,  {J{\o}rgensen} J.~K.,  {van Dishoeck} E.~F.,   {de Wachter}
  E.~B.~M.,  2007, \mn@doi [\aap] {10.1051/0004-6361:20065963}, \href
  {http://adsabs.harvard.edu/abs/2007A%26A...465..913B} {465, 913}

\bibitem[\protect\citeauthoryear{{Bisschop}, {J{\o}rgensen}, {Bourke},
  {Bottinelli}  \& {van Dishoeck}}{{Bisschop} et~al.}{2008}]{bisschop2008}
{Bisschop} S.~E.,  {J{\o}rgensen} J.~K.,  {Bourke} T.~L.,  {Bottinelli} S.,
  {van Dishoeck} E.~F.,  2008, \mn@doi [\aap] {10.1051/0004-6361:200809673},
  \href {http://adsabs.harvard.edu/abs/2008A%26A...488..959B} {488, 959}

\bibitem[\protect\citeauthoryear{{Blake}, {Sutton}, {Masson}  \&
  {Phillips}}{{Blake} et~al.}{1987}]{blake1987}
{Blake} G.~A.,  {Sutton} E.~C.,  {Masson} C.~R.,   {Phillips} T.~G.,  1987,
  \mn@doi [\apj] {10.1086/165165}, \href
  {http://adsabs.harvard.edu/abs/1987ApJ...315..621B} {315, 621}

\bibitem[\protect\citeauthoryear{{Boogert}, {Gerakines}  \&
  {Whittet}}{{Boogert} et~al.}{2015}]{boogert2015}
{Boogert} A.~C.~A.,  {Gerakines} P.~A.,   {Whittet} D.~C.~B.,  2015, \mn@doi
  [\araa] {10.1146/annurev-astro-082214-122348}, \href
  {http://adsabs.harvard.edu/abs/2015ARA%26A..53..541B} {53, 541}

\bibitem[\protect\citeauthoryear{{Bossa}, {Paardekooper}, {Isokoski}  \&
  {Linnartz}}{{Bossa} et~al.}{2015}]{bossa2015}
{Bossa} J.-B.,  {Paardekooper} D.~M.,  {Isokoski} K.,   {Linnartz} H.,  2015,
  \mn@doi [Physical Chemistry Chemical Physics (Incorporating Faraday
  Transactions)] {10.1039/C5CP00578G}, \href
  {http://adsabs.harvard.edu/abs/2015PCCP...1717346B} {17, 17346}

\bibitem[\protect\citeauthoryear{{Bottinelli} et~al.,}{{Bottinelli}
  et~al.}{2004}]{bottinelli2004}
{Bottinelli} S.,  et~al., 2004, \mn@doi [\apjl] {10.1086/426964}, \href
  {http://adsabs.harvard.edu/abs/2004ApJ...617L..69B} {617, L69}

\bibitem[\protect\citeauthoryear{{Cazaux}, {Tielens}, {Ceccarelli}, {Castets},
  {Wakelam}, {Caux}, {Parise}  \& {Teyssier}}{{Cazaux}
  et~al.}{2003}]{cazaux2003}
{Cazaux} S.,  {Tielens} A.~G.~G.~M.,  {Ceccarelli} C.,  {Castets} A.,
  {Wakelam} V.,  {Caux} E.,  {Parise} B.,   {Teyssier} D.,  2003, \mn@doi
  [\apjl] {10.1086/378038}, \href
  {http://adsabs.harvard.edu/abs/2003ApJ...593L..51C} {593, L51}

\bibitem[\protect\citeauthoryear{{Cernicharo} et~al.,}{{Cernicharo}
  et~al.}{2016}]{cernicharo2016}
{Cernicharo} J.,  et~al., 2016, \mn@doi [\aap] {10.1051/0004-6361/201527531},
  \href {http://adsabs.harvard.edu/abs/2016A%26A...587L...4C} {587, L4}

\bibitem[\protect\citeauthoryear{{Chen}, {Chuang}, {Mu{\~n}oz Caro}, {Nuevo},
  {Chu}, {Yih}, {Ip}  \& {Wu}}{{Chen} et~al.}{2014}]{chen2014}
{Chen} Y.-J.,  {Chuang} K.-J.,  {Mu{\~n}oz Caro} G.~M.,  {Nuevo} M.,  {Chu}
  C.-C.,  {Yih} T.-S.,  {Ip} W.-H.,   {Wu} C.-Y.~R.,  2014, \mn@doi [\apj]
  {10.1088/0004-637X/781/1/15}, \href
  {http://adsabs.harvard.edu/abs/2014ApJ...781...15C} {781, 15}

\bibitem[\protect\citeauthoryear{{Coutens} et~al.,}{{Coutens}
  et~al.}{2016}]{coutens2016}
{Coutens} A.,  et~al., 2016, \mn@doi [\aap] {10.1051/0004-6361/201628612},
  \href {http://adsabs.harvard.edu/abs/2016A%26A...590L...6C} {590, L6}

\bibitem[\protect\citeauthoryear{{Crockett} et~al.,}{{Crockett}
  et~al.}{2014}]{crockett2014}
{Crockett} N.~R.,  et~al., 2014, \mn@doi [\apj] {10.1088/0004-637X/787/2/112},
  \href {http://adsabs.harvard.edu/abs/2014ApJ...787..112C} {787, 112}

\bibitem[\protect\citeauthoryear{{Fedoseev}, {Ioppolo}, {Zhao}, {Lamberts}  \&
  {Linnartz}}{{Fedoseev} et~al.}{2015}]{fedoseev2015}
{Fedoseev} G.,  {Ioppolo} S.,  {Zhao} D.,  {Lamberts} T.,   {Linnartz} H.,
  2015, \mn@doi [\mnras] {10.1093/mnras/stu2028}, \href
  {http://adsabs.harvard.edu/abs/2015MNRAS.446..439F} {446, 439}

\bibitem[\protect\citeauthoryear{{Fedoseev}, {Chuang}, {van Dishoeck},
  {Ioppolo}  \& {Linnartz}}{{Fedoseev} et~al.}{2016}]{fedoseev2016}
{Fedoseev} G.,  {Chuang} K.-J.,  {van Dishoeck} E.~F.,  {Ioppolo} S.,
  {Linnartz} H.,  2016, \mn@doi [\mnras] {10.1093/mnras/stw1028}, \href
  {http://adsabs.harvard.edu/abs/2016MNRAS.460.4297F} {460, 4297}

\bibitem[\protect\citeauthoryear{{Feuchtgruber}, {Helmich}, {van Dishoeck}  \&
  {Wright}}{{Feuchtgruber} et~al.}{2000}]{feuchtgruber2000}
{Feuchtgruber} H.,  {Helmich} F.~P.,  {van Dishoeck} E.~F.,   {Wright} C.~M.,
  2000, \mn@doi [\apjl] {10.1086/312711}, \href
  {http://adsabs.harvard.edu/abs/2000ApJ...535L.111F} {535, L111}

\bibitem[\protect\citeauthoryear{{Garrod}}{{Garrod}}{2013}]{garrod2013}
{Garrod} R.~T.,  2013, \mn@doi [\apj] {10.1088/0004-637X/778/2/158}, \href
  {http://adsabs.harvard.edu/abs/2013ApJ...778..158G} {778, 158}

\bibitem[\protect\citeauthoryear{{Goesmann} et~al.,}{{Goesmann}
  et~al.}{2015}]{goesmann2015}
{Goesmann} F.,  et~al., 2015, \mn@doi [Science] {10.1126/science.aab0689},
  \href {http://adsabs.harvard.edu/abs/2015Sci...349b0689G} {349}

\bibitem[\protect\citeauthoryear{{Halfen}, {Ilyushin}  \& {Ziurys}}{{Halfen}
  et~al.}{2015}]{halfen2015}
{Halfen} D.~T.,  {Ilyushin} V.~V.,   {Ziurys} L.~M.,  2015, \mn@doi [\apjl]
  {10.1088/2041-8205/812/1/L5}, \href
  {http://adsabs.harvard.edu/abs/2015ApJ...812L...5H} {812, L5}

\bibitem[\protect\citeauthoryear{{Henderson} \& {Gudipati}}{{Henderson} \&
  {Gudipati}}{2015}]{hendersongudipati2015}
{Henderson} B.~L.,  {Gudipati} M.~S.,  2015, \mn@doi [\apj]
  {10.1088/0004-637X/800/1/66}, \href
  {http://adsabs.harvard.edu/abs/2015ApJ...800...66H} {800, 66}

\bibitem[\protect\citeauthoryear{{Herbst} \& {van Dishoeck}}{{Herbst} \& {van
  Dishoeck}}{2009}]{herbstdishoeck2009}
{Herbst} E.,  {van Dishoeck} E.~F.,  2009, \mn@doi [\araa]
  {10.1146/annurev-astro-082708-101654}, \href
  {http://adsabs.harvard.edu/abs/2009ARA%26A..47..427H} {47, 427}

\bibitem[\protect\citeauthoryear{{Jaber}, {Ceccarelli}, {Kahane}  \&
  {Caux}}{{Jaber} et~al.}{2014}]{jaber2014}
{Jaber} A.~A.,  {Ceccarelli} C.,  {Kahane} C.,   {Caux} E.,  2014, \mn@doi
  [\apj] {10.1088/0004-637X/791/1/29}, \href
  {http://adsabs.harvard.edu/abs/2014ApJ...791...29J} {791, 29}

\bibitem[\protect\citeauthoryear{{J{\o}rgensen}, {Bourke}, {Nguyen Luong}  \&
  {Takakuwa}}{{J{\o}rgensen} et~al.}{2011}]{jorgensen2011}
{J{\o}rgensen} J.~K.,  {Bourke} T.~L.,  {Nguyen Luong} Q.,   {Takakuwa} S.,
  2011, \mn@doi [\aap] {10.1051/0004-6361/201117139}, \href
  {http://adsabs.harvard.edu/abs/2011A%26A...534A.100J} {534, A100}

\bibitem[\protect\citeauthoryear{{J{\o}rgensen}, {Favre}, {Bisschop}, {Bourke},
  {van Dishoeck}  \& {Schmalzl}}{{J{\o}rgensen} et~al.}{2012}]{jorgensen2012}
{J{\o}rgensen} J.~K.,  {Favre} C.,  {Bisschop} S.~E.,  {Bourke} T.~L.,  {van
  Dishoeck} E.~F.,   {Schmalzl} M.,  2012, \mn@doi [\apjl]
  {10.1088/2041-8205/757/1/L4}, \href
  {http://adsabs.harvard.edu/abs/2012ApJ...757L...4J} {757, L4}

\bibitem[\protect\citeauthoryear{{J{\o}rgensen} et~al.,}{{J{\o}rgensen}
  et~al.}{2016}]{jorgensen2016}
{J{\o}rgensen} J.~K.,  et~al., 2016, \mn@doi [\aap]
  {10.1051/0004-6361/201628648}, \href
  {http://adsabs.harvard.edu/abs/2016A%26A...595A.117J} {595, A117}

\bibitem[\protect\citeauthoryear{{Kolesnikov{\'a}}, {Alonso}, {Berm{\'u}dez},
  {Alonso}, {Tercero}, {Cernicharo}  \& {Guillemin}}{{Kolesnikov{\'a}}
  et~al.}{2016}]{kolesnikova2016}
{Kolesnikov{\'a}} L.,  {Alonso} J.~L.,  {Berm{\'u}dez} C.,  {Alonso} E.~R.,
  {Tercero} B.,  {Cernicharo} J.,   {Guillemin} J.-C.,  2016, \mn@doi [\aap]
  {10.1051/0004-6361/201628140}, \href
  {http://adsabs.harvard.edu/abs/2016A%26A...591A..75K} {591, A75}

\bibitem[\protect\citeauthoryear{{Koput}}{{Koput}}{1986}]{koput1986}
{Koput} J.,  1986, \mn@doi [Journal of Molecular Spectroscopy]
  {10.1016/0022-2852(86)90281-X}, \href
  {http://adsabs.harvard.edu/abs/1986JMoSp.115..131K} {115, 131}

\bibitem[\protect\citeauthoryear{{Kuan} et~al.,}{{Kuan}
  et~al.}{2004}]{kuan2004}
{Kuan} Y.-J.,  et~al., 2004, \mn@doi [\apjl] {10.1086/426315}, \href
  {http://adsabs.harvard.edu/abs/2004ApJ...616L..27K} {616, L27}

\bibitem[\protect\citeauthoryear{{Ligterink}, {Paardekooper}, {Chuang}, {Both},
  {Cruz-Diaz}, {van Helden}  \& {Linnartz}}{{Ligterink}
  et~al.}{2015}]{ligterink2015}
{Ligterink} N.~F.~W.,  {Paardekooper} D.~M.,  {Chuang} K.-J.,  {Both} M.~L.,
  {Cruz-Diaz} G.~A.,  {van Helden} J.~H.,   {Linnartz} H.,  2015, \mn@doi
  [\aap] {10.1051/0004-6361/201526930}, \href
  {http://adsabs.harvard.edu/abs/2015A%26A...584A..56L} {584, A56}

\bibitem[\protect\citeauthoryear{Linnartz, Ioppolo  \& Fedoseev}{Linnartz
  et~al.}{2015}]{linnartz2015}
Linnartz H.,  Ioppolo S.,   Fedoseev G.,  2015, \mn@doi [International Reviews
  in Physical Chemistry] {10.1080/0144235X.2015.1046679}, 34, 205

\bibitem[\protect\citeauthoryear{{L{\'o}pez-Sepulcre}
  et~al.,}{{L{\'o}pez-Sepulcre} et~al.}{2015}]{lopez-sepulcre2015}
{L{\'o}pez-Sepulcre} A.,  et~al., 2015, \mn@doi [\mnras]
  {10.1093/mnras/stv377}, \href
  {http://adsabs.harvard.edu/abs/2015MNRAS.449.2438L} {449, 2438}

\bibitem[\protect\citeauthoryear{{Lykke} et~al.,}{{Lykke}
  et~al.}{2017}]{lykke2017}
{Lykke} J.~M.,  et~al., 2017, \mn@doi [\aap] {10.1051/0004-6361/201629180},
  \href {http://adsabs.harvard.edu/abs/2017A%26A...597A..53L} {597, A53}

\bibitem[\protect\citeauthoryear{{Mart{\'{\i}}n-Dom{\'e}nech}, {Rivilla},
  {Jim{\'e}nez-Serra}, {Quenard}, {Testi}  \&
  {Mart{\'{\i}}n-Pintado}}{{Mart{\'{\i}}n-Dom{\'e}nech}
  et~al.}{2017}]{martin-domenech2017}
{Mart{\'{\i}}n-Dom{\'e}nech} R.,  {Rivilla} V.,  {Jim{\'e}nez-Serra} I.,
  {Quenard} D.,  {Testi} L.,   {Mart{\'{\i}}n-Pintado} J.,  2017, preprint,
  \href {http://adsabs.harvard.edu/abs/2017arXiv170104376M} {} (\mn@eprint
  {arXiv} {1701.04376})

\bibitem[\protect\citeauthoryear{{Milam}, {Savage}, {Brewster}, {Ziurys}  \&
  {Wyckoff}}{{Milam} et~al.}{2005}]{milam2005}
{Milam} S.~N.,  {Savage} C.,  {Brewster} M.~A.,  {Ziurys} L.~M.,   {Wyckoff}
  S.,  2005, \mn@doi [\apj] {10.1086/497123}, \href
  {http://adsabs.harvard.edu/abs/2005ApJ...634.1126M} {634, 1126}

\bibitem[\protect\citeauthoryear{{M{\"u}ller}, {Thorwirth}, {Roth}  \&
  {Winnewisser}}{{M{\"u}ller} et~al.}{2001}]{muller2001}
{M{\"u}ller} H.~S.~P.,  {Thorwirth} S.,  {Roth} D.~A.,   {Winnewisser} G.,
  2001, \mn@doi [\aap] {10.1051/0004-6361:20010367}, \href
  {http://adsabs.harvard.edu/abs/2001A%26A...370L..49M} {370, L49}

\bibitem[\protect\citeauthoryear{{M{\"u}ller}, {Schl{\"o}der}, {Stutzki}  \&
  {Winnewisser}}{{M{\"u}ller} et~al.}{2005}]{muller2005}
{M{\"u}ller} H.~S.~P.,  {Schl{\"o}der} F.,  {Stutzki} J.,   {Winnewisser} G.,
  2005, \mn@doi [Journal of Molecular Structure]
  {10.1016/j.molstruc.2005.01.027}, \href
  {http://adsabs.harvard.edu/abs/2005JMoSt.742..215M} {742, 215}

\bibitem[\protect\citeauthoryear{{Neill} et~al.,}{{Neill}
  et~al.}{2014}]{neill2014}
{Neill} J.~L.,  et~al., 2014, \mn@doi [\apj] {10.1088/0004-637X/789/1/8}, \href
  {http://adsabs.harvard.edu/abs/2014ApJ...789....8N} {789, 8}

\bibitem[\protect\citeauthoryear{{Nummelin}, {Bergman}, {Hjalmarson},
  {Friberg}, {Irvine}, {Millar}, {Ohishi}  \& {Saito}}{{Nummelin}
  et~al.}{2000}]{nummelin2000}
{Nummelin} A.,  {Bergman} P.,  {Hjalmarson} {\AA}.,  {Friberg} P.,  {Irvine}
  W.~M.,  {Millar} T.~J.,  {Ohishi} M.,   {Saito} S.,  2000, \mn@doi [\apjs]
  {10.1086/313376}, \href {http://adsabs.harvard.edu/abs/2000ApJS..128..213N}
  {128, 213}

\bibitem[\protect\citeauthoryear{{{\"O}berg}, {Garrod}, {van Dishoeck}  \&
  {Linnartz}}{{{\"O}berg} et~al.}{2009}]{oberg2009a}
{{\"O}berg} K.~I.,  {Garrod} R.~T.,  {van Dishoeck} E.~F.,   {Linnartz} H.,
  2009, \mn@doi [\aap] {10.1051/0004-6361/200912559}, \href
  {http://adsabs.harvard.edu/abs/2009A%26A...504..891O} {504, 891}

\bibitem[\protect\citeauthoryear{{{\"O}berg}, {Boogert}, {Pontoppidan}, {van
  den Broek}, {van Dishoeck}, {Bottinelli}, {Blake}  \& {Evans}}{{{\"O}berg}
  et~al.}{2011}]{oberg2011}
{{\"O}berg} K.~I.,  {Boogert} A.~C.~A.,  {Pontoppidan} K.~M.,  {van den Broek}
  S.,  {van Dishoeck} E.~F.,  {Bottinelli} S.,  {Blake} G.~A.,   {Evans} II
  N.~J.,  2011, \mn@doi [\apj] {10.1088/0004-637X/740/2/109}, \href
  {http://adsabs.harvard.edu/abs/2011ApJ...740..109O} {740, 109}

\bibitem[\protect\citeauthoryear{{Parise}, {Simon}, {Caux}, {Dartois},
  {Ceccarelli}, {Rayner}  \& {Tielens}}{{Parise} et~al.}{2003}]{parise2003}
{Parise} B.,  {Simon} T.,  {Caux} E.,  {Dartois} E.,  {Ceccarelli} C.,
  {Rayner} J.,   {Tielens} A.~G.~G.~M.,  2003, \mn@doi [\aap]
  {10.1051/0004-6361:20031277}, \href
  {http://adsabs.harvard.edu/abs/2003A%26A...410..897P} {410, 897}

\bibitem[\protect\citeauthoryear{{Pineda} et~al.,}{{Pineda}
  et~al.}{2012}]{pineda2012}
{Pineda} J.~E.,  et~al., 2012, \mn@doi [\aap] {10.1051/0004-6361/201219589},
  \href {http://adsabs.harvard.edu/abs/2012A%26A...544L...7P} {544, L7}

\bibitem[\protect\citeauthoryear{{Raunier}, {Chiavassa}, {Duvernay}, {Borget},
  {Aycard}, {Dartois}  \& {d'Hendecourt}}{{Raunier} et~al.}{2004}]{raunier2004}
{Raunier} S.,  {Chiavassa} T.,  {Duvernay} F.,  {Borget} F.,  {Aycard} J.~P.,
  {Dartois} E.,   {d'Hendecourt} L.,  2004, \mn@doi [\aap]
  {10.1051/0004-6361:20034558}, \href
  {http://adsabs.harvard.edu/abs/2004A%26A...416..165R} {416, 165}

\bibitem[\protect\citeauthoryear{{Reva}, {Lapinski}  \& {Fausto}}{{Reva}
  et~al.}{2010}]{reva2010}
{Reva} I.,  {Lapinski} L.,   {Fausto} R.,  2010, \mn@doi [J. Mol. Struct.]
  {10.1016/j.molstruc.2010.03.081}, 976, 333

\bibitem[\protect\citeauthoryear{{Ruzi} \& {Anderson}}{{Ruzi} \&
  {Anderson}}{2012}]{ruzi2012}
{Ruzi} M.,  {Anderson} D.,  2012, \mn@doi [J. Chem. Phys.] {10.1063/1.4765372},
  137, 194313

\bibitem[\protect\citeauthoryear{{Sakaizumi}, {Mure}, {Ohashi}  \&
  {Yamaguchi}}{{Sakaizumi} et~al.}{1990}]{sakaizumi1990}
{Sakaizumi} T.,  {Mure} H.,  {Ohashi} O.,   {Yamaguchi} I.,  1990, \mn@doi
  [Journal of Molecular Spectroscopy] {10.1016/0022-2852(90)90007-D}, \href
  {http://adsabs.harvard.edu/abs/1990JMoSp.140...62S} {140, 62}

\bibitem[\protect\citeauthoryear{{Sullivan}, {Heusel}, {Zunic}  \&
  {Durig}}{{Sullivan} et~al.}{1994}]{sullivan1994}
{Sullivan} J.,  {Heusel} H.,  {Zunic} W.,   {Durig} J.,  1994, Spectrochimica
  Acta, 50

\bibitem[\protect\citeauthoryear{{Teolis}, {Loeffler}, {Raut}, {Fam{\'a}}  \&
  {Baragiola}}{{Teolis} et~al.}{2007}]{teolis2007}
{Teolis} B.~D.,  {Loeffler} M.~J.,  {Raut} U.,  {Fam{\'a}} M.,   {Baragiola}
  R.~A.,  2007, \mn@doi [\icarus] {10.1016/j.icarus.2007.03.023}, \href
  {http://adsabs.harvard.edu/abs/2007Icar..190..274T} {190, 274}

\bibitem[\protect\citeauthoryear{{Tercero}, {Kleiner}, {Cernicharo}, {Nguyen},
  {L{\'o}pez}  \& {Mu{\~n}oz Caro}}{{Tercero} et~al.}{2013}]{tercero2013}
{Tercero} B.,  {Kleiner} I.,  {Cernicharo} J.,  {Nguyen} H.~V.~L.,  {L{\'o}pez}
  A.,   {Mu{\~n}oz Caro} G.~M.,  2013, \mn@doi [\apjl]
  {10.1088/2041-8205/770/1/L13}, \href
  {http://adsabs.harvard.edu/abs/2013ApJ...770L..13T} {770, L13}

\bibitem[\protect\citeauthoryear{{Wilson} \& {Rood}}{{Wilson} \&
  {Rood}}{1994}]{wilson1994}
{Wilson} T.~L.,  {Rood} R.,  1994, \mn@doi [\araa]
  {10.1146/annurev.aa.32.090194.001203}, \href
  {http://adsabs.harvard.edu/abs/1994ARA%26A..32..191W} {32, 191}

\bibitem[\protect\citeauthoryear{{Winnewisser}, {Pearson}, {Galica}  \&
  {Winnewisser}}{{Winnewisser} et~al.}{1982}]{winnewisser1982}
{Winnewisser} M.,  {Pearson} E.~F.,  {Galica} J.,   {Winnewisser} B.~P.,  1982,
  \mn@doi [Journal of Molecular Spectroscopy] {10.1016/0022-2852(82)90044-3},
  \href {http://adsabs.harvard.edu/abs/1982JMoSp..91..255W} {91, 255}

\bibitem[\protect\citeauthoryear{{Zhou} \& {Durig}}{{Zhou} \&
  {Durig}}{2009}]{zhou2009}
{Zhou} S.,  {Durig} J.,  2009, JMS, 924

\bibitem[\protect\citeauthoryear{{van Broekhuizen}, {Keane}  \& {Schutte}}{{van
  Broekhuizen} et~al.}{2004}]{vanbroekhuizen2004}
{van Broekhuizen} F.~A.,  {Keane} J.~V.,   {Schutte} W.~A.,  2004, \mn@doi
  [\aap] {10.1051/0004-6361:20034161}, \href
  {http://adsabs.harvard.edu/abs/2004A%26A...415..425V} {415, 425}

\bibitem[\protect\citeauthoryear{{van Broekhuizen}, {Pontoppidan}, {Fraser}  \&
  {van Dishoeck}}{{van Broekhuizen} et~al.}{2005}]{vanbroekhuizen2005}
{van Broekhuizen} F.~A.,  {Pontoppidan} K.~M.,  {Fraser} H.~J.,   {van
  Dishoeck} E.~F.,  2005, \mn@doi [\aap] {10.1051/0004-6361:20041711}, \href
  {http://adsabs.harvard.edu/abs/2005A%26A...441..249V} {441, 249}

\bibitem[\protect\citeauthoryear{{van Dishoeck}, {Blake}, {Jansen}  \&
  {Groesbeck}}{{van Dishoeck} et~al.}{1995}]{vandishoeck1995}
{van Dishoeck} E.~F.,  {Blake} G.~A.,  {Jansen} D.~J.,   {Groesbeck} T.~D.,
  1995, \mn@doi [\apj] {10.1086/175915}, \href
  {http://adsabs.harvard.edu/abs/1995ApJ...447..760V} {447, 760}

\makeatother
\end{thebibliography}



\appendix

\section{Line list}

Table \ref{table_obs} contains the full list of detected CH$_{3}$NCO, v$_b$=0 transitions toward IRAS16293. A total of 43 lines are observed.

\begin{table*}
\caption{Unblended lines of CH$_3$NCO detected toward source B.} 
\begin{center}
\begin{tabular}{lccccccc}
\hline \hline
Species & Transition & Frequency & $E_{\rm up}$ & $A_{\rm ij}$ & $g_{\rm up}$ & \\
 & & (MHz) & (K) & (s$^{-1}$) & & \\
 \hline
CH$_3$NCO, v$_b$=0 & (38 0 0 2 -- 37 0 0 2) & 329675.6 & 361.3 & 1.71\,$\times$\,10$^{-3}$ & 77 \\
CH$_3$NCO, v$_b$=0 & (38 1 0 2 -- 37 1 0 2) & 329732.1 & 367.3 & 1.71\,$\times$\,10$^{-3}$ & 77 \\
CH$_3$NCO, v$_b$=0 & (38 -1 0 -3 -- 37 -1 0 -3) & 329929.3 & 427.1 & 1.67\,$\times$\,10$^{-3}$ & 77 \\
CH$_3$NCO, v$_b$=0 & (38 -1 0 2 -- 37 -1 0 2) & 330611.8 & 367.3 & 1.71\,$\times$\,10$^{-3}$ & 77 \\
CH$_3$NCO, v$_b$=0 & (38 1 0 3 -- 37 1 0 3) & 331251.7 & 431.6 & 1.74\,$\times$\,10$^{-3}$ & 77 \\
CH$_3$NCO, v$_b$=0 & (38 0 0 -3 -- 37 0 0 -3) & 331337.3 & 423.4 & 1.70\,$\times$\,10$^{-3}$ & 77\\
CH$_3$NCO, v$_b$=0 & (38 2 0 3 -- 37 2 0 3) & 331653.6 & 447.5 & 1.72\,$\times$\,10$^{-3}$ & 77 \\
CH$_3$NCO, v$_b$=0 & (38 3 0 -3 -- 37 3 0 -3) & 332324.9 & 477.7 & 1.71\,$\times$\,10$^{-3}$ & 77 \\
CH$_3$NCO, v$_b$=0 & (39 0 0 1 -- 38 0 0 1) & 334214.7 & 336.5 & 1.82\,$\times$\,10$^{-3}$ & 79 \\
CH$_3$NCO, v$_b$=0 & (39 1 39 0 -- 38 1 38 0) & 334470.6 & 327.3 & 1.78\,$\times$\,10$^{-3}$ & 79 \\
CH$_3$NCO, v$_b$=0 & (39 -1 0 1 -- 38 -1 0 1) & 335432.3 & 342.5 & 1.83\,$\times$\,10$^{-3}$ & 79 \\
CH$_3$NCO, v$_b$=0 & (39 2 0 2 -- 38 2 0 2) & 335521.3 & 401.3 & 1.83\,$\times$\,10$^{-3}$ & 79 \\
CH$_3$NCO, v$_b$=0 & (39 3 0 2 -- 38 3 0 2) & 335973.3 & 431.1 & 1.82\,$\times$\,10$^{-3}$ & 79 \\
CH$_3$NCO, v$_b$=0 & (39 0 39 0 -- 38 0 38 0) & 336339.9 & 323.7 & 1.82\,$\times$\,10$^{-3}$ & 79 \\
CH$_3$NCO, v$_b$=0 & (39 2 38 0 -- 38 2 37 0) & 337737.4 & 348.4 & 1.83\,$\times$\,10$^{-3}$ & 79 \\
CH$_3$NCO, v$_b$=0 & (39 0 0 2 -- 38 0 0 2) & 338235.6 & 377.6 & 1.85\,$\times$\,10$^{-3}$ & 79 \\
CH$_3$NCO, v$_b$=0 & (39 2 37 0 -- 38 2 36 0) & 339028.2 & 349.0 & 1.85\,$\times$\,10$^{-3}$ & 79 \\
CH$_3$NCO, v$_b$=0 & (38 3 0 1 -- 37 3 0 1) & 339370.5 & 374.5 & 1.75\,$\times$\,10$^{-3}$ & 77 \\
CH$_3$NCO, v$_b$=0 & (39 1 0 -3 -- 38 1 0 -3) & 339848.8 & 448.8 & 1.88\,$\times$\,10$^{-3}$ & 79 \\
CH$_3$NCO, v$_b$=0 & (39 1 0 3 -- 38 1 0 3) & 339948.0 & 448.0 & 1.88\,$\times$\,10$^{-3}$ & 79 \\
CH$_3$NCO, v$_b$=0 & (39 1 38 0 -- 38 1 37 0) & 340327.9 & 333.1 & 1.88\,$\times$\,10$^{-3}$ & 79 \\
CH$_3$NCO, v$_b$=0 & (39 2 0 -3 -- 38 2 0 -3) & 340363.3 & 464.6 & 1.87\,$\times$\,10$^{-3}$ & 79 \\
CH$_3$NCO, v$_b$=0 & (40 0 0 1 -- 39 0 0 1) & 342747.6 & 353.1 & 1.97\,$\times$\,10$^{-3}$ & 81 \\
CH$_3$NCO, v$_b$=0 & (40 1 0 1 -- 39 1 0 1) & 344280.7 & 359.2 & 1.98\,$\times$\,10$^{-3}$ & 81 \\
CH$_3$NCO, v$_b$=0 & (40 3 0 2 -- 39 3 0 2)  & 344650.3 & 447.7 & 1.97\,$\times$\,10$^{-3}$ & 81 \\
CH$_3$NCO, v$_b$=0 & (40 3 38 0 -- 39 3 37 0) & 344660.0 & 395.0 & 1.97\,$\times$\,10$^{-3}$ & 81 \\
CH$_3$NCO, v$_b$=0 & (40 -1 0 -3 -- 39 -1 0 -3) & 346771.8 & 459.6 & 1.95\,$\times$\,10$^{-3}$ & 81 \\
CH$_3$NCO, v$_b$=0 & (40 0 0 2 -- 39 0 0 2) & 346789.2 & 394.2 & 1.99\,$\times$\,10$^{-3}$ & 81 \\
CH$_3$NCO, v$_b$=0 & (40 1 0 2 -- 39 1 0 2) & 346813.1 & 400.2 & 1.99\,$\times$\,10$^{-3}$ & 81 \\
CH$_3$NCO, v$_b$=0 & (40 1 0 -3 -- 39 1 0 -3) & 348541.5 & 465.6 & 2.03\,$\times$\,10$^{-3}$ & 81 \\
CH$_3$NCO, v$_b$=0 & (40 0 0 -3 -- 39 0 0 -3) & 348768.0 & 456.2 & 1.99\,$\times$\,10$^{-3}$ & 81 \\
CH$_3$NCO, v$_b$=0 & (40 -3 0 2 -- 39 -3 0 2) & 348867.2 & 447.9 & 1.99\,$\times$\,10$^{-3}$ & 81 \\
CH$_3$NCO, v$_b$=0 & (41 1 0 1 -- 40 1 0 1) & 352855.7 & 376.2 & 2.13\,$\times$\,10$^{-3}$ & 83 \\
CH$_3$NCO, v$_b$=0 & (41 3 0 2 -- 40 3 0 2) & 353345.6 & 464.8 & 2.12\,$\times$\,10$^{-3}$ & 83 \\
CH$_3$NCO, v$_b$=0 & (41 -1 0 2 -- 40 -1 0 2) & 356670.2 & 417.3 & 2.15\,$\times$\,10$^{-3}$ & 83 \\
CH$_3$NCO, v$_b$=0 & (41 1 40 0 -- 40 1 39 0) & 357668.8 & 367.0 & 2.18\,$\times$\,10$^{-3}$ & 83 \\
CH$_3$NCO, v$_b$=0 & (41 2 0 -3 -- 40 2 0 -3) & 357766.6 & 498.4 & 2.17\,$\times$\,10$^{-3}$ & 83 \\
CH$_3$NCO, v$_b$=0 & (41 0 0 5 -- 40 0 0 5) & 358444.3 & 671.4 & 2.16\,$\times$\,10$^{-3}$ & 83 \\
CH$_3$NCO, v$_b$=0 & (42 1 42 0 -- 41 1 41 0) & 360081.7 & 378.0 & 2.23\,$\times$\,10$^{-3}$ & 85 \\
CH$_3$NCO, v$_b$=0 & (42 2 0 2 -- 41 2 0 2) & 361199.5 & 452.4 & 2.28\,$\times$\,10$^{-3}$ & 85 \\
CH$_3$NCO, v$_b$=0 & (42 1 0 1 -- 41 1 0 1) & 361425.5 & 393.7 & 2.29\,$\times$\,10$^{-3}$ & 85 \\
CH$_3$NCO, v$_b$=0 & (42 3 40 0 -- 41 3 39 0) & 361770.5 & 429.5 & 2.28\,$\times$\,10$^{-3}$ & 85 \\
CH$_3$NCO, v$_b$=0 & (42 3 0 2 -- 41 3 0 2) & 362063.6 & 482.2 & 2.28\,$\times$\,10$^{-3}$ & 85 \\
\hline
\end{tabular}
\end{center}
\label{table_obs}
\end{table*}

\section{CH$_{3}$NCO v$_{b}$=1}

\begin{figure*}
 \centering
  \includegraphics[width=\hsize]{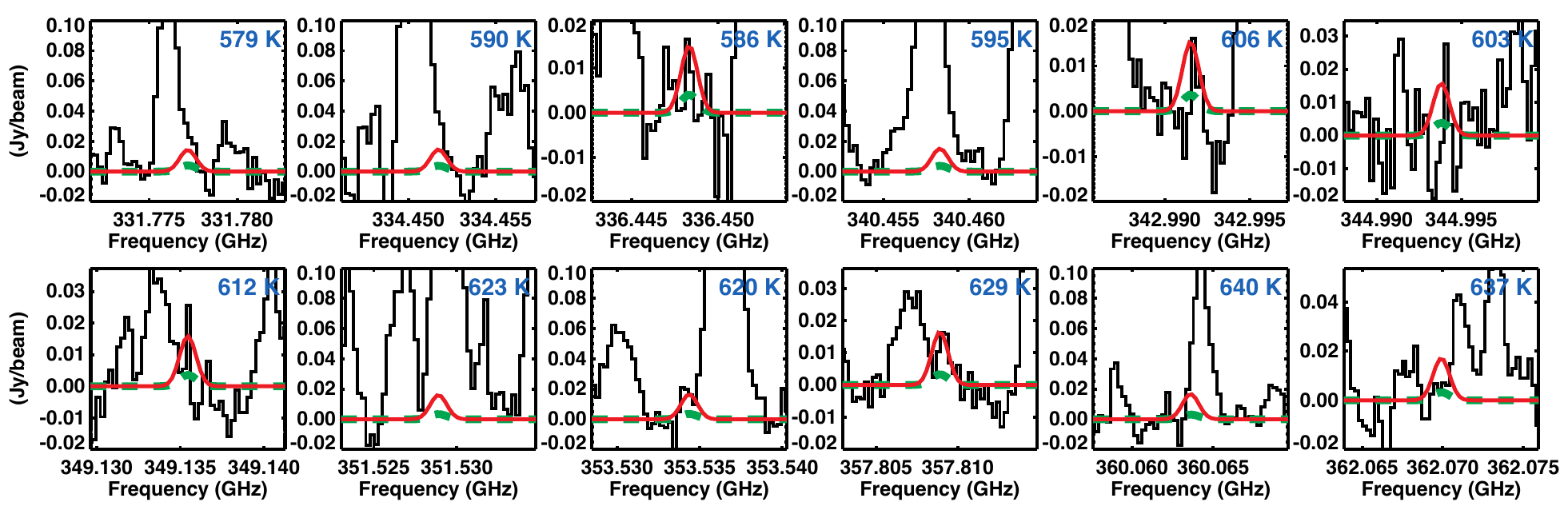}
 \caption{Predictions for the CH$_3$NCO $v_b$=1 lines toward source B (red solid: best-fit model for $T_{\rm ex}$ = 300 K, green dashed: best-fit model for $T_{\rm ex}$ = 100 K).}
 \label{fig.ch3nco_vb1}
\end{figure*}

Figure \ref{fig.ch3nco_vb1} shows the tentative identification of the first excited state of methyl isocyanate. 

\section{$m/z$ = 56, 57 and 58 upon VUV irradiation of pure CH$_{4}$ and HNCO samples}
\label{sec.mass_CH4_HNCO}

\begin{figure}
\begin{center}
\includegraphics[width=\hsize]{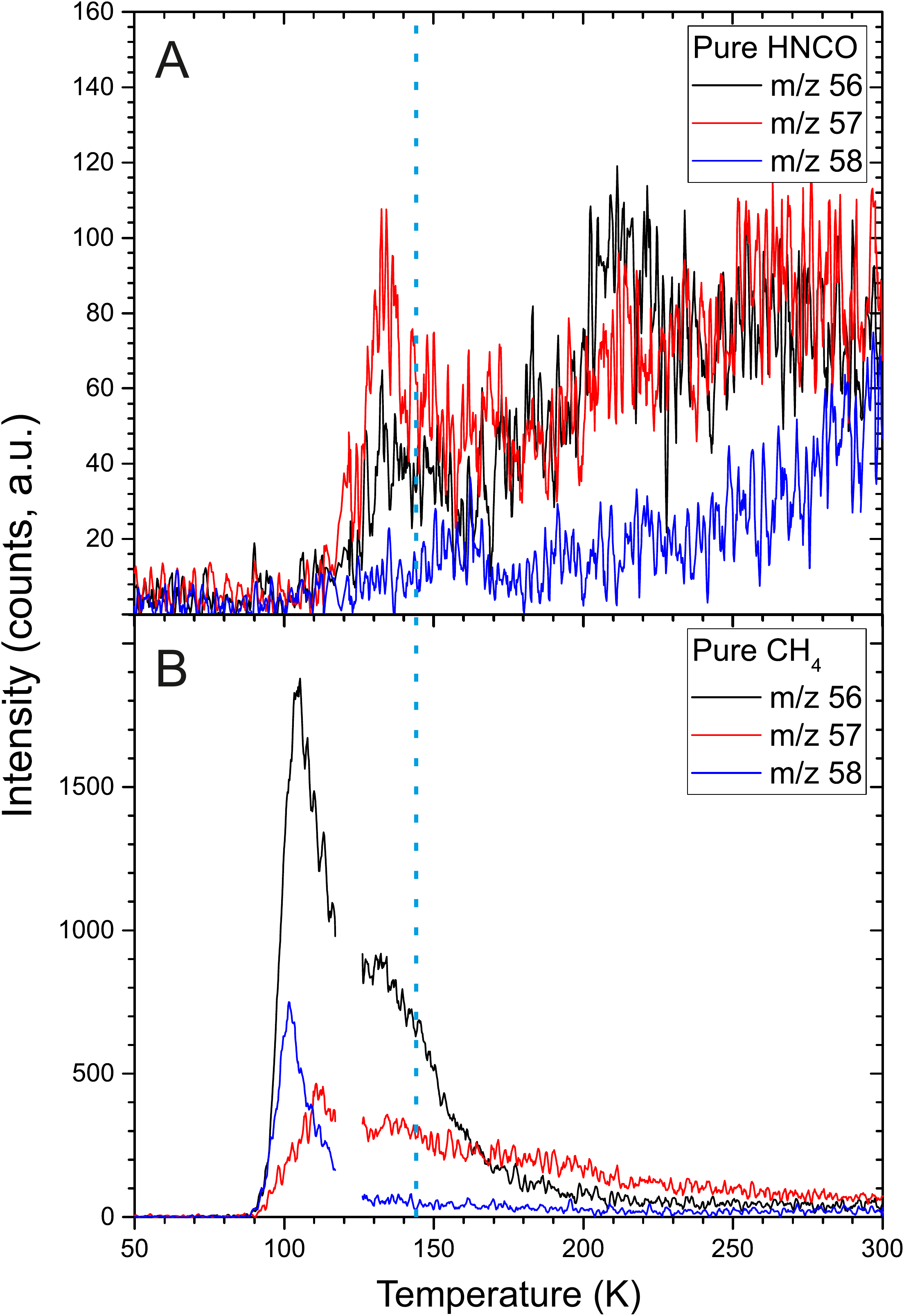}
\caption{TPD trace of $m/z$ = 57 (primary mass CH$_{3}$NCO), $m/z$ = 56 (secondary mass CH$_{3}$NCO) and $m/z$ = 58 after the irradiation of pure HNCO (A) and pure CH$_{4}$ (B). The desorption peak found in the CH$_{4}$:HNCO mixtures is indicated by the dashed line at 145 K.}
\label{fig.mass_CH4_HNCO}
\end{center}
\end{figure}

Several publications address the energetic processing of pure HNCO and CH$_{4}$ ice and much is known about the chemistry that can be induced in these species \citep{raunier2004,bennett2006,bossa2015}. Of the known chemical products, none of them contributes to the primary and secondary masses of methyl isocyanate. To verify this, pure methane and isocyanic acid have been VUV irradiated. The HNCO sample (Figure \ref{fig.mass_CH4_HNCO}A) does not show any significant release of these masses. However, after irradiation of methane a substantial amount of $m/z$ = 56, coinciding with a lesser amount of $m/z$ = 58, is seen releasing at a desorption temperature of 105 K. Some $m/z$ 57 is seen to be released as well, although there is no clear peak found in this case. The resulting contribution of $m/z$ = 56 can interfere with the $m/z$ = 57/56 amu mass ratio of methyl isocyanate.

\bsp	
\label{lastpage}
\end{document}